\documentclass[reprint,amsmath,amssymb,aps,pra,showkeys]{revtex4-2}

\usepackage{graphicx}
\usepackage{bm,color}
\usepackage{booktabs}
\usepackage{array}
\usepackage{braket}

\allowdisplaybreaks

\newcolumntype{L}[1]{>{\raggedright\let\newline\\\arraybackslash\hspace{0pt}}m{#1}}

\renewcommand{\d}{\text{d}}
\renewcommand{\i}{\text{i}}
\newcommand{\e}{\text{e}}

\newcommand{\ind}[1]{_{\text{#1}}}
\newcommand{\GM}{\mathbf{G}^{\text{M}}}
\newcommand{\aM}{\mathbf{a}^{\text{M}}}
\newcommand{\R}{\mathbf{R}}

\newcommand{\Q}{\mathbf{Q}}
\renewcommand{\r}{\mathbf{r}}
\newcommand{\q}{\mathbf{q}}
\newcommand{\tr}{\tilde{\r}}
\newcommand{\eps}{\epsilon_\mathrm{r}\epsilon_0}
\newcommand{\er}{\epsilon_\mathrm{r}}
\newcommand{\kb}{k_\mathrm{B}}

\begin{document}

\title{Moiré-Bose-Hubbard model for interlayer excitons in twisted transition metal dichalcogenide heterostructures}%
\author{Niclas Götting}
\author{Frederik Lohof}
\author{Christopher Gies}
\affiliation{Institute for Theoretical Physics, University of Bremen, Bremen, Germany
}%

\date{\today}

\begin{abstract}
In bilayers of semiconducting transition metal dichalcogenides, the twist angle between layers can be used to introduce a highly regular periodic potential modulation on a length scale that is large compared to the unit cell. In such structures, correlated states can emerge, in which excitons in the heterostructure are strongly localized to the potential minima due to exciton-exciton interactions. We explore the transition between Mott and extended exciton phases in terms of a moiré-Bose-Hubbard Hamiltonian. Hopping and on-site interaction parameters are obtained from a Wannier representation of the interlayer-exciton wave functions, and a non-local Rytova-Keldysh model is used to attribute for the dielectric screening of excitons in the two-dimensional material. For sufficiently small exciton concentrations and substrate screening our model predicts the emergence of Mott-insulating states, establishing twisted TMD heterostructures as possible quantum simulators for bosonic many-body systems.
\end{abstract}


\maketitle

\section{Introduction}

\begin{figure}[t]
    \centering
    \includegraphics[width=\linewidth]{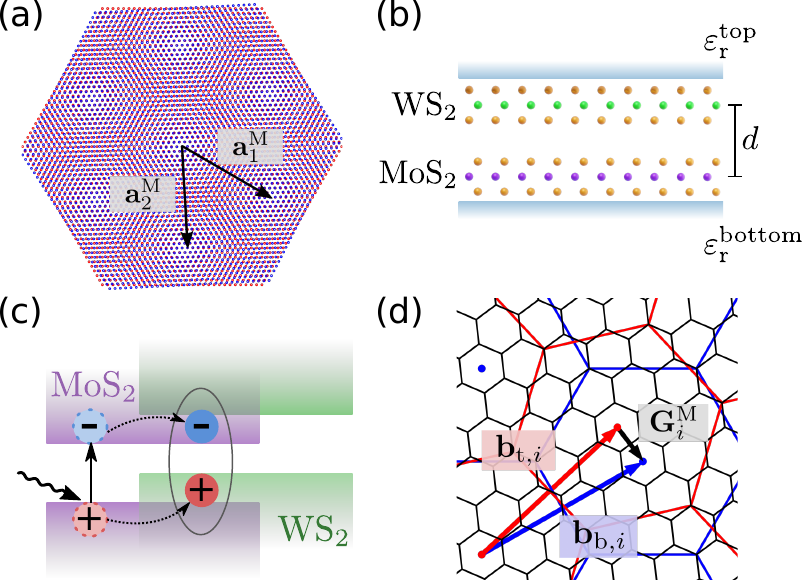}
    \caption{(a) Real space lattices of top and bottom monolayers. The periodic variation of the stacking registry defines the moiré lattice vectors $\aM_i$ that span the moiré unit cell. (b) Illustration of the moiré heterostructure. The two constituting TMD monolayers are embedded between top and bottom dielectric materials. (c) Schematic arrangement of the monolayer band gaps to form a type-II band alignment. Generation of electron-hole pairs, fast interlayer charge transfer, and IX formation are indicated. (d) A relative twist of the monolayer Brillouin zones allows the construction of a subordinate moiré lattice and corresponding unit cell. The moiré lattice vectors $\GM_i = \mathbf{b}_{\mathrm{b}, i} - \mathbf{b}_{\mathrm{t}, i}$ are generated from differences of lattice vectors of the top and bottom layers.}
    \label{fig:geometry}
\end{figure}

Twisted heterobilayers of transition metal dichalcogenides (TMDs) offer unique opportunities to study many-body and quantum phenomena, such as strongly correlated states of matter and unusual topological phases \cite{wilson_excitons_2021, kennes_moire_2021}. The twist angle between two adjacent layers of a heterostructure gives rise to periodically changing stacking order -- a so-called moiré pattern (Fig.~\ref{fig:geometry}(a)) -- resulting in a regular variation of the band gap due to different interlayer hybridization strengths \cite{wu_topological_2017,wu_theory_2018,tran_evidence_2019,brem_tunable_2020}. Charge carriers and excitonic complexes confined to the bilayer material experience an effective potential with a periodicity of the arising superlattice, which, for small twist angles, is much larger than the unit cell size of the individual monolayers. Crucially, the moiré superlattice defines a periodic array of potential minima, in which charge carriers can be trapped and interact with each other -- a behavior reminiscent of Hubbard systems \cite{hubbard_electron_1963,fisher_boson_1989}. Only recently, studies on gated van der Waals heterostructures have demonstrated the extent of this analogy by revealing correlated hole-Mott phases at integer and fractional filling factors of charge carriers, i.e.~a realization of an effective Fermi-Hubbard system  \cite{tang_simulation_2020, huang_correlated_2021}. Theoretical models have also predicted topological phases, charge density waves, and generalized Wigner crystals \cite{wu_hubbard_2018, slagle_charge_2020, pan_band_2020, pan_quantum_2020}.

While the Fermi-Hubbard physics of twisted moiré heterostructures is still in its infancy, even less studied is the Bose-Hubbard physics of moiré excitons and other bosonic complexes. In TMD heterobilayers with type-II band alignment, intra- and interlayer excitons can exist \cite{torun_interlayer_2018,jiang_interlayer_2021}. We focus on interlayer excitons (IX), which are energetically favorable and dominate the emission properties. They exhibit strong dipole alignment due to the separation of electrons and holes in the two constituting planes of the bilayer material. The strong Coulomb interaction in atomically thin materials gives reason to treat IXs as composite bosonic particles, so that their mutual interaction in the presence of the regular moiré potential realizes a generalized Bose-Hubbard (BH) system. On the one hand, TMD moiré systems may thus serve as a promising semiconductor platform for simulating many-body effects and quantum phase transitions in addition to dilute ultracold atomic gases in optical lattices \cite{greiner_quantum_2002,jaksch_cold_2005}. On the other hand, predictions of the moiré-Bose-Hubbard model can provide insight into the widely unexplored phase boundaries of IX. First studies have already investigated the possibility of superfluid phase formation in TMD heterobilayers \cite{lagoin_key_2021, wang_diffusivity_2021}, exploring the transition between superfluid and Mott state. Realizations of BH systems have also been studied with IX in gated double quantum-well structures \cite{lagoin_microscopic_2020, lagoin_quasicondensation_2021}, where the lattice potential was realized by patterned, externally applied electric fields. In comparison, in twisted TMD heterobilayers the moiré potential emerges naturally from the varying stacking order in the twisted material layers, where the degree of localization can be altered via the twist angle.

In the present work, we start from an effective description of IX in twisted heterobilayers and calculate the characteristic on-site and hopping parameters directly from the IX wave functions in a localized Wannier basis. It is well known that excitons in atomically thin van der Waals layers are highly susceptible to screening by the dielectric environment \cite{steinke_noninvasive_2017, borghardt_engineering_2017, kylanpaa_binding_2015, berman_superfluidity_2017}. We account for dielectric screening by using a non-local Rytova-Keldysh potential \cite{rytova1967the8248, keldysh_coulomb_1979, cudazzo_dielectric_2011, rodin_excitons_2014} and explore the effect of different dielectric environments on the phase boundary between the Mott and superfluid state of IXs in twisted TMD heterobilayers for the typical material combinations MoS$_2$/WS$_2$ and MoSe$_2$/WSe$_2$ \cite{terrones_novel_2013, wu_theory_2018, wu_topological_2017,jiang_interlayer_2021, torun_interlayer_2018}. We begin by discussing the emergence of the moiré potential and its influence on the energy structure of IXs in Section \ref{sec:moire_IX}, while in Section \ref{sec:moire_BH} we introduce a single-band Wannier basis to map the IX's dynamics and their dipolar interaction onto a generalized BH model, extracting hopping amplitudes as well as on-site and nearest-neighbor interaction parameters. Section \ref{sec:mod_screening} lays out the model used to account for screening of the dipolar interaction due to the dielectric environment. Finally, in Section \ref{sec:discussion} we discuss implications for IX phases that arise in MoS$_2$/WS$_2$ and MoSe$_2$/WSe$_2$ moiré-Bose-Hubbard systems.

\section{Interlayer Excitons in a Moiré superlattice}
\label{sec:moire_IX}
We first focus on twisted heterobilayers of MoS$_2$/WS$_2$, embedded in a dielectric environment, as depicted in Fig.~\ref{fig:geometry}(b). For negligible mismatch in lattice constants, the emerging moiré lattice is only determined by the twist angle. The moiré lattice constant can be approximated by $|\aM|\approx a_0/\theta$, with $a_0$ the lattice constant of the TMD monolayers. The spatial twist of the constituent monolayers translates to an offset of the K points in reciprocal space. It shifts the K-valley extrema of each respective monolayer relative to each other to form a type-II band alignment that is indirect in momentum space with the conduction-band minimum in the MoS$_2$ and the valence-band maximum in the WS$_2$ layer as seen in Fig.~\ref{fig:geometry}(c). As illustrated in Fig.~\ref{fig:geometry}(d) (black), an effective smaller moiré Brillouin zone (MBZ) emerges that is due to the long-range periodicity of the real lattice. Outer corners of the MBZ are denoted by $\kappa$. The resulting moiré lattice vectors are given by the differences of the reciprocal lattice vectors of the top (t) and bottom (b) monolayers $\GM_i = \mathbf{b}_{\mathrm{b}, i} - \mathbf{b}_{\mathrm{t}, i}$ with $(\mathbf{b}_{\mathrm{t}/\mathrm{b}, 1}, \mathbf{b}_{\mathrm{t}/\mathrm{b}, 2}) = 2\pi \left( (\mathbf{a}_{\mathrm{t}/\mathrm{b}, 1}, \mathbf{a}_{\mathrm{t}/\mathrm{b}, 2})^\intercal \right)^{-1}$ and $\mathbf{a}_{\mathrm{t}/\mathrm{b}, i}$ the real space lattice vectors.

Electron-hole pairs generated in the heterostructure quickly separate due to the type-II band alignment and fast carrier scattering \cite{hong_ultrafast_2014, chen_ultrafast_2016,jin_ultrafast_2018}, leaving electrons residing in the conduction band of the MoS$_2$ layer and holes in the valence band of the WS$_2$ layer. In the presence of the strong Coulomb interaction, momentum-indirect IXs form that possess long radiative lifetimes \cite{choi_twist_2021}, allowing them to thermalize before recombining. Furthermore, with charge carriers separated to different layers, the IX dipole moments align perpendicular to the in-plane direction, an arrangement that is in close analogy to spatially-indirect excitons in coupled quantum wells, and which results in a pronounced dipolar repulsion \cite{zimmermann_excitonexciton_2007, schindler_analysis_2008, laikhtman_exciton_2009}.

\begin{figure}
  \centering
  \includegraphics[width=\linewidth]{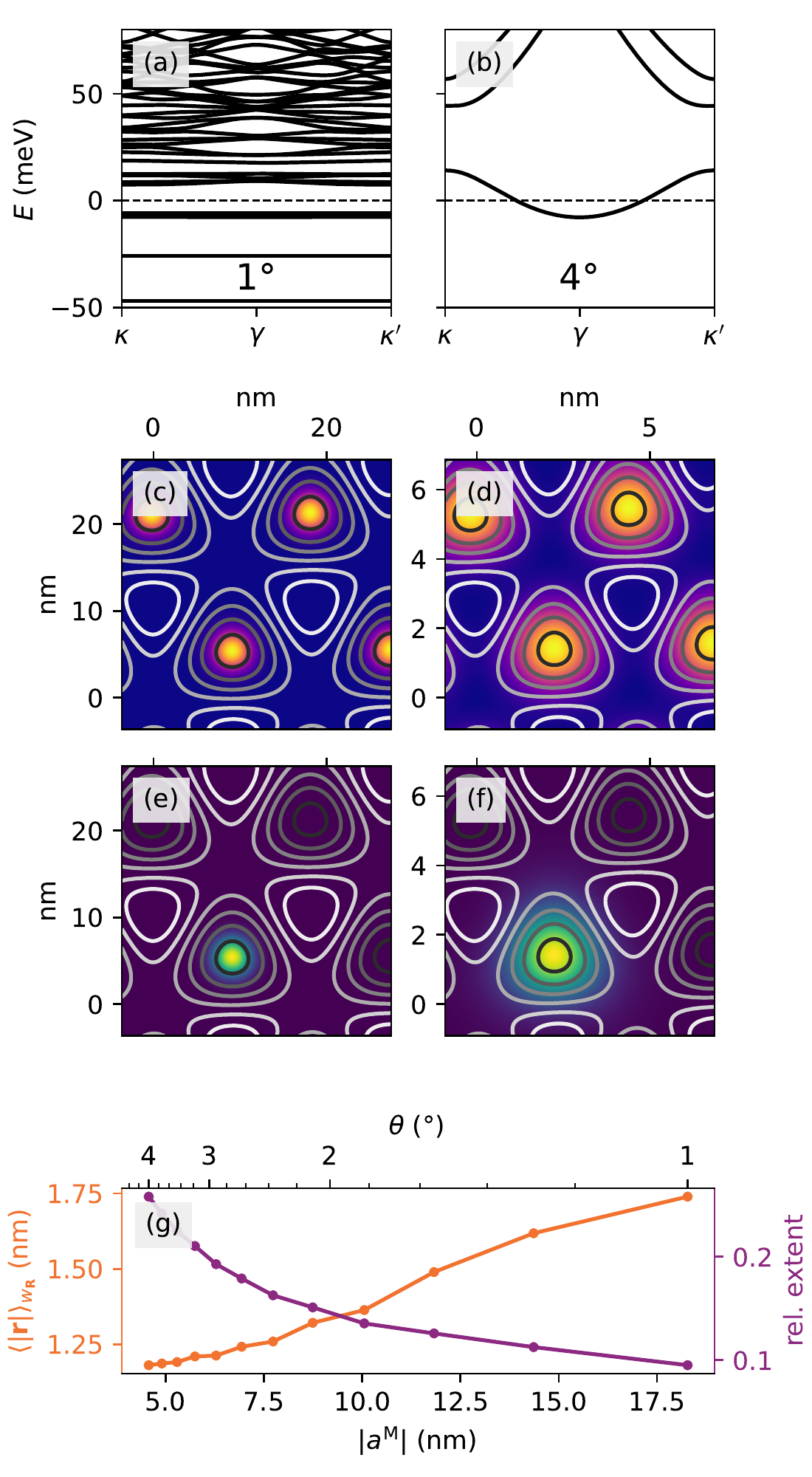}
  \caption{(a), (b) IX dispersion at twist angles $\theta=1^\circ$ and $4^\circ$ along a linear cut through the MBZ ($\kappa, \gamma, \kappa'$). Smaller twist angles lead to stronger localization of the IX wave functions at lowest energies (flat dispersion). (c)--(d) IX Bloch-wave functions (colors). The wave function's localization is reflected in the spatial extent relative to the moiré potential (contours). (e)--(f) Single-band IX Wannier functions are localized at individual potential minima. (g) First absolute moment $\langle |\r|\rangle_{w_{\R}}$ and relative extent as a function of moiré lattice length and twist angle $\theta$. }
  \label{fig:wave_functions}
\end{figure}

For small twist angles, the periodic, continuous spatial variation of the interlayer band gap acts as an effective potential landscape $V^\mathrm{M}(\r)$ for the IXs. Although also the binding energy varies with the moiré periodicity, this variation is small, and we neglect it here \cite{wu_theory_2018}. The effective potential generally modifies the movement of IXs and can give rise to states localized at the potential minima when the potential is sufficiently strong.
In a first-order plane-wave expansion, the potential can be written as \cite{wu_theory_2018}
\begin{equation}
 V^\mathrm{M}(\r) = \sum_{j=1}^6 V_j \, e^{i\GM_j \r} ,
\end{equation}
where $\bm{G}_j$ are the first-shell moiré reciprocal lattice vectors. Because of the threefold rotational symmetry of the heterobilayer $V_1=V_3=V_5$, $V_2=V_4=V_6$, and $V_1 = V_2^\ast$. Using the Bohr radius of intralayer excitons as a rough estimate, we note that for twist angles $< 4^\circ$, the moiré periodicity is sufficiently large in comparison to the in-plane extent of IXs \cite{wu_theory_2018}. Therefore, in the limit of small densities, we regard the IX as composite bosonic particles that move in the moiré potential landscape, neglecting their inner structure. Considering only the center-of-mass (c.o.m.) kinetic energy of an IX and the influence of the moiré potential, the Hamiltonian reads
\begin{equation}
  H_0 = -\frac{\hbar^2}{2M}\Delta_{\r} + V^\mathrm{M}(\r)
\end{equation}
where $M = m_\mathrm{e} + m_\mathrm{h}$ is the total mass of the IX
which we solve in a plane-wave basis of the IX Bloch states. The resulting IX dispersion for twist angles 1° and 4° is shown in Fig.~\ref{fig:wave_functions}(a)--(b) and parameters used are listed in Table~\ref{tab:parameters}. The strong localization of the IX is reflected in the vanishing curvature of the lowest IX energy bands. The resulting IX Bloch functions for states with in-plane quasimomentum $\Q$ in the branch $\alpha$ of the dispersion are given by
\begin{equation}
    \chi_{\Q}^{(\alpha)}(\r) = \frac{1}{\sqrt{V}}\sum_{\GM}c^{(\alpha)}_{\Q - \GM}\e^{\i(\Q - \GM)\r}.
\end{equation}
\begin{table}
\caption{Parameters of two material systems for the calculation of the IX dispersion and wave functions. Total IX mass $M_\mathrm{IX} = m_\mathrm{e} + m_\mathrm{h}$ is the sum of electron and hole masses given in units of the free electron mass $m_0$. Lattice constants $a_0$ of the respective monolayers (assumed equal for top and bottom layer) and interlayer distance $d$ are in units of nm. For stacking configurations AA the parameters are given for the moiré potential according to $V = |V|\e^{i\psi}$ with $V$ in units of meV and $\psi$ in degree (°). Parameters are taken from \cite{wu_theory_2018}.}
\begin{ruledtabular}
\begin{tabular}{L{2cm}rrrrrr}
Material combination & $M_\mathrm{IX}$ & $a_0$ & $d$ & Stacking & $|V|$ & $\psi$ \\
\cmidrule{1-4}
\cmidrule{5-7}
MoS$_2$/WS$_2$ 		& 0.76 	& 0.319	 & 0.615	& AA & 12.4 & 81.5 \\
& \\
MoSe$_2$/WSe$_2$ 	& 0.84 	& 0.332	 & 0.647 & AA & 11.8 & 79.5 \\
\end{tabular}
\end{ruledtabular}
\label{tab:parameters}
\end{table}In Fig.~\ref{fig:wave_functions}(c)--(d) we show the moiré potential (contours) superimposed over the IX Bloch wave functions (colors) corresponding to the lowest energy states at the $\kappa$ point of the moiré BZ. While still periodic in real space, the wave functions become more localized around the potential minima as the dispersion flattens with decreasing twist angle.

The IXs feel an interaction potential \cite{batsch_dipole-dipole_1993, li_dipolar_2020} that can be obtained from a point-charge treatment \cite{schindler_analysis_2008} of two spatially-indirect excitons. The resulting repulsive dipole-dipole interaction is given by
\begin{equation}
  U(r) = \frac{e^2}{4\pi\eps} \left(\frac{2}{r} - \frac{2}{\sqrt{d^2 + r^2}}\right)~, \label{eq:U_r}
\end{equation}
where $d$ is the mean distance between the layers (i.e. the length of the dipole). Due to its 2D nature, in the heterobilayer the effect of dielectric screening is particularly important as it strongly modifies the Coulomb interaction between particles in the layers \cite{steinke_noninvasive_2017, borghardt_engineering_2017, kylanpaa_binding_2015, berman_superfluidity_2017}. The simplest method to treat this effect is to account for the constant relative permittivity $\epsilon_\mathrm{r}$ as the (equally) weighted average of the relative permittivities (see Table \ref{tab:screenings}) of the two constituting monolayers and the top and bottom sub- and superstrates. An improved model of the of dielectric screening by the heterobilayer and its environment is discussed in Section \ref{sec:mod_screening}.

Given the dipolar repulsion, we estimate under which conditions one can regard the IXs as composite bosonic particles. We assume that this assumption holds as long as the mean IX distance $\bar{r} = n^{-1/2}$ is much larger than the interlayer distance $d$, given an IX density $n$. The distance $d$ defines the characteristic length scale of the potential, beyond which the IXs see each other as point-like dipoles. This results in the condition $nd^2\ll 1$, which in our case amounts to $n \ll 2\times 10^{14}\,\mathrm{cm}^{-2}$ for the MoS$_2$/WS$_2$ heterostructure, and smaller densities if additional dielectric screening layers are placed between the TMD monolayers. In the limit of freely moving  particles, we need sufficiently low temperatures, such that the minimal distance that IXs can have due mean kinetic energy is large, and they do not witness each other's internal structure. This is provided if the characteristic length scale $d$ of the dipolar potential is small in comparison to the minimal distance $r_E$ of two IXs with thermal energy $E = \kb T$, where  $r_E$ is determined by $U(r_E) = E$. From this we obtain $\kb T \ll \frac{e^2}{4\pi \eps d}$, which is met for all experimental conditions. In the case of IXs confined to minima of the moiré potential, we consider a density regime that results in at most one IX per potential minimum. This leads to densities $n < \frac{2}{\sqrt{3}|\aM|}$ as a function of the length of the moiré lattice vector, see also Appendix \ref{sec:density_estimates} for more details.

\section{Moiré Bose-Hubbard Model.}
\label{sec:moire_BH}

\begin{figure}
  \centering
  \includegraphics[width=\linewidth]{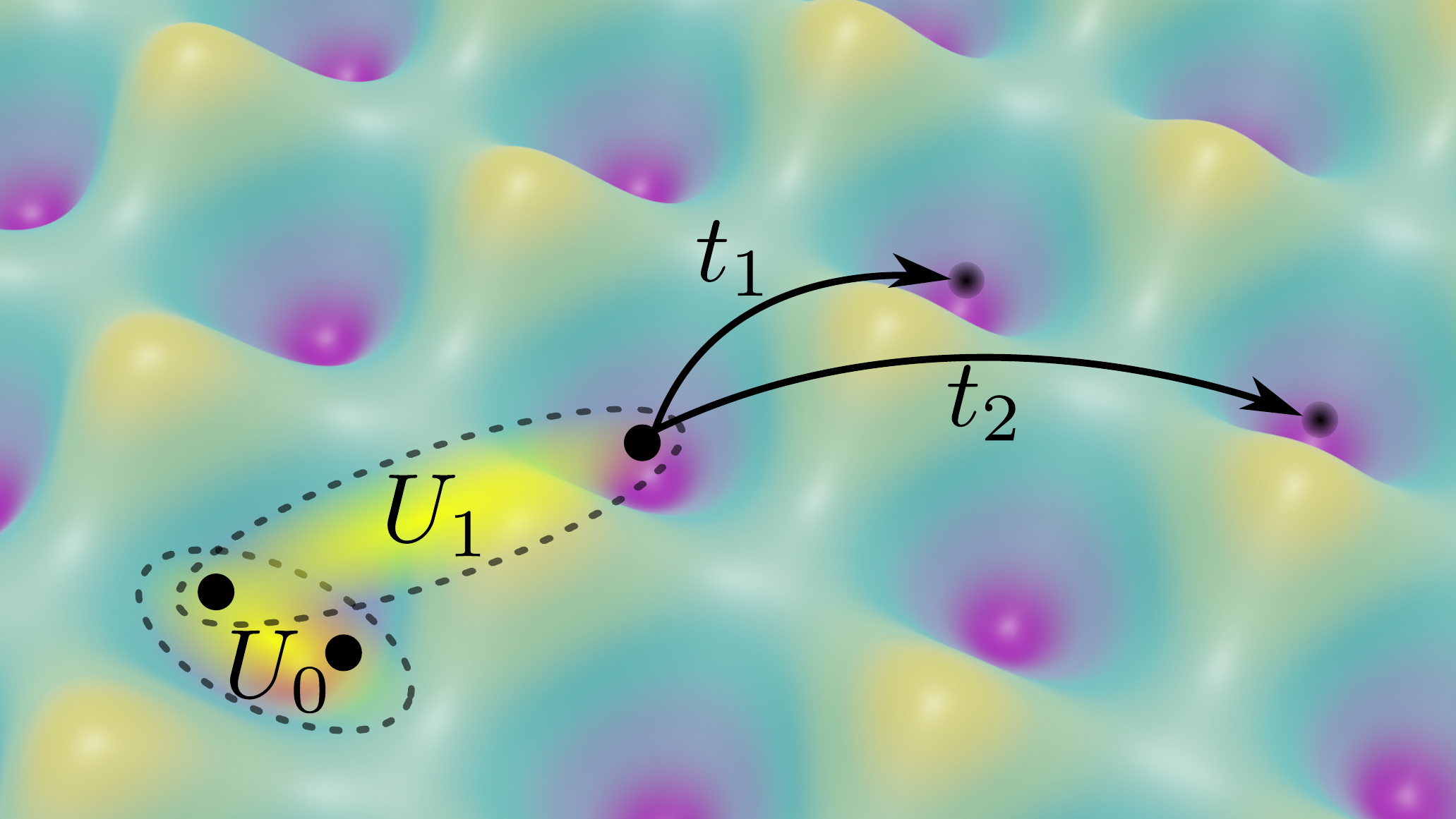}
  \caption{Schematic view of the interaction processes represented by the Hubbard parameters. On-site and nearest-neighbor interaction strengths $U_0$ and $U_1$ are indicated as well as hopping to nearest ($t_1$) and next-nearest ($t_2$) lattice sites.}
  \label{fig:BH_illustration}
\end{figure}

\begin{figure*}
    \centering
    \includegraphics[width=\linewidth]{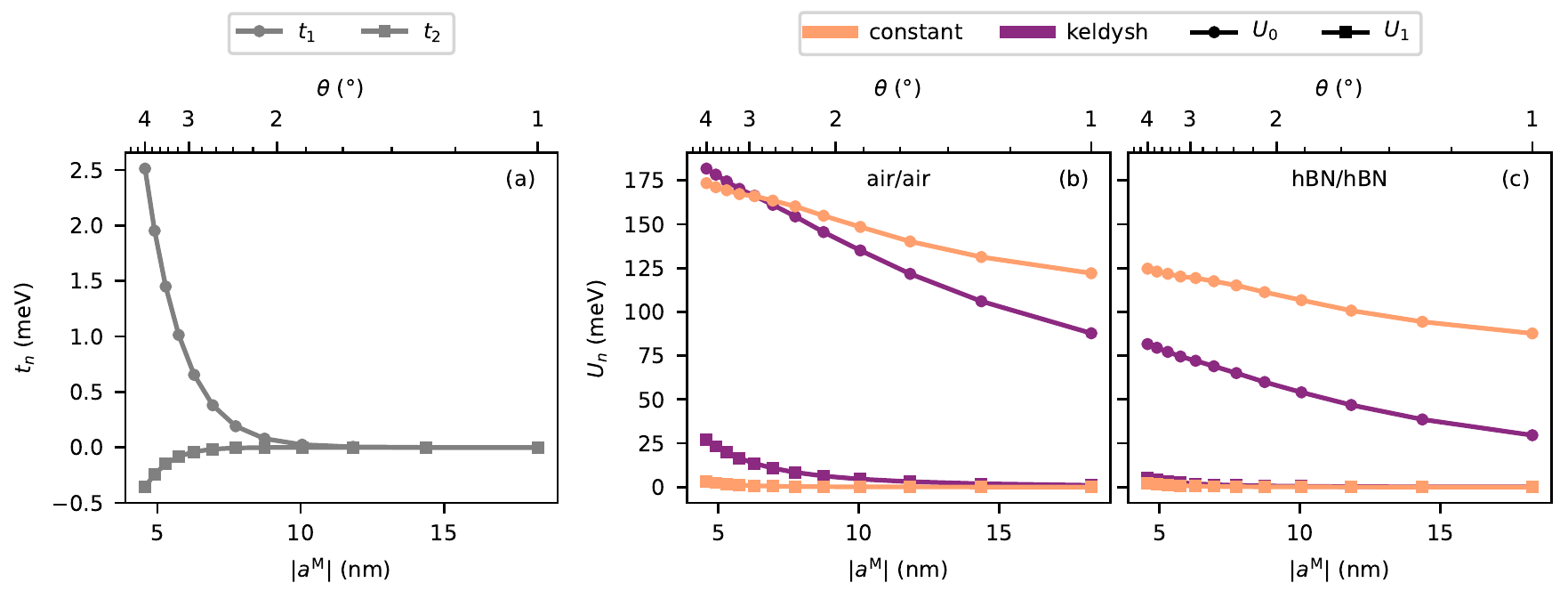}
    \caption{Moiré Bose-Hubbard parameters. (a) The (next-) nearest-neighbor hopping amplitudes $t_1$, $t_2$. Increasing localization of the IX wave functions with decreasing twist angle $\theta$ result in suppressed hopping. (b)--(c) On-site and nearest-neighbor interaction $U_0$ (circles) and $U_1$ (squares). Constant screening (orange) is compared with screening given by the non-local Keldysh potential (purple). For both dielectric embeddings (freestanding, hBN/hBN) the Keldysh model predicts weaker on-site interactions $U_0$ (except in the freestanding case for larger angles) and an increased role of nearest-neighbor interactions $U_1$. With decreasing twist angles, the dipolar interaction weakens as the moiré unit cell becomes large and IXs occupy a larger area in absolute terms. }
    \label{fig:t_U}
\end{figure*}

The localization of lowest-energy moiré IXs in the effective potential minima allows to map the system to an extended Bose-Hubbard model on a triangular lattice. We use a basis of Wannier functions, constructed from the moiré IX wave functions in the lowest band
\begin{equation}
    w_{\R}(\r) = \frac{1}{\sqrt{N}}\sum_{\Q}\e^{-\i\Q\R}\chi_\Q(\r), \label{eq:wannier_def}
\end{equation}
where $\chi_\Q(\r)$ are the IX c.o.m.\ wave functions for in-plane momenta $\Q$ in the two-dimensional MBZ, and from here on we suppress the band index $\alpha$. Note that the $\R$ do not correspond to the moiré lattice vectors but to the array of moiré potential minima around which the Wannier functions $w_{\R}(\r)$ are centered. The applicability of single-band Wannier functions relies on the energetic separation of the lowest IX state as seen in Fig.~\ref{fig:wave_functions}(a)--(b). This can be assumed to be a good approximation for sufficiently low exciton densities, where higher energy bands are largely empty. The $w_{\R}(\r)$ are centered around individual minima $\R$ of the moiré potential as shown in Fig.~\ref{fig:wave_functions}(e)--(f). The ability to map the system dynamics to that of a Bose-Hubbard model relies on choosing the $w_{\R}(\r)$ as basis functions that are strongly localized at discrete lattice sites. For an intuitive way to quantify the localization of the IXs we calculate the first absolute moment of the distribution defined by $w_{\R}(\r)$ centered around its mean:
\begin{equation}
\langle |\r|\rangle_{w_{\R}} = \int  |\r-\R| |w_{\R}(\r)|^2 \mathrm{d}^2 r\label{eq:abs_extent}
\end{equation}
As seen in Fig.~\ref{fig:wave_functions}(g), $\langle |\r|\rangle_{w_{\R}}$ actually increases with decreasing twist angle. Therefore, we note that localization of IX states at lattice sites occurs only relative to the extent of the moiré unit cell, which can be inferred from the same panel, in which also the \emph{relative} extent $\langle |\r|\rangle_{w_{\R}} / |\aM|$ is shown, $|\aM|$ being the moiré lattice vector defining the length scale of the moiré lattice.

Having established the c.o.m.\ Hamiltonian of the IXs and their dipolar interaction in the previous sections, we now seek a description of IX states in the TMD moiré heterostructure in terms of a generalized Bose-Hubbard model, which includes nearest-neighbor IX repulsion to account for the long-range interaction mediated by the dipolar potential $U(r)$. In general, we write the BH-Hamiltonian
\begin{equation}
    H = \sum_{\R,\R'} t(\R -\R') c_\R^\dag c_{\R'} + \frac{1}{2} \sum_{\R,\R'} U(\R -\R') c_\R^\dag c_{\R'}^\dag c_{\R'} c_{\R} \label{eq:bh}
\end{equation}
with $c_\R^\dag$, $ c_\R$ the creation and annihilation operators for IXs at lattice position $\R$. The hopping amplitudes $t_n = -t(\R -\R')$ and the interaction strengths $U_n = U(\R -\R')$ are obtained by projecting the free IX Hamiltonian and the potential $U(r)$ onto the first-band Wannier basis states in equation \eqref{eq:wannier_def}. Here, a two-center approximation was applied that relies on the spatially fast decaying basis functions $w_{\R}(\r)$. With this notation, $U_0$ corresponds to on-site interaction with $\R =\R'$, while $U_1$ denotes nearest-neighbor interaction. Generally, $U_n$ is the $n^{\mathrm{th}}$-nearest-neighbor interaction, and similarly for $t_n$. We define the hopping parameters $t_n$ with a negative sign in accordance with the literature on BH systems. A sketch illustrating the relevant processes is shown in Fig.~\ref{fig:BH_illustration}.
To obtain $t_n$ we use the projection of $H_0$ onto the Wannier basis

\begin{equation}
    t_n = -\int_{\mathbb{R}^2} w^*_{\R}(\r) H_0 w_{\R'}(\r)\d^2 r = -\frac{1}{N} \sum_{\Q}\e^{\i(\R - \R')\Q} E_{\Q},\label{eq:tn}
\end{equation}
where $E_{\Q}$ are the eigenvalues of $H_0$ for the lowest dispersion branch. We note that $t_0$ establishes a constant energy offset given by the mean energy of the band $E_{\Q}$ that we neglect in the following. The algebraic form of the $t_n$ parameter gives interesting insight: Effectively, $t_n$ is a Fourier decomposition of the lowest energy band with respect to the lattice site distance $\R - \R'$. Information about the parameter can be directly read off of the band structure, i.e.~flat bands result in small hopping terms, while curved bands result in larger hopping terms. We show $t_n$ in Fig.~\ref{fig:t_U}(a) as a function of twist angle. Indeed, for small twist angles the localization of IXs becomes stronger, effectively inhibiting hopping from one lattice site to another.

Similarly to $t_n$, the coefficients of the two-particle interaction $U_n$ can be calculated using the Wannier basis in second quantization using Eq.~\eqref{eq:U_r} as
\begin{equation}
    U_n = \iint_{\mathbb{R}^2} |w_{\R}(\r)|^2 |w_{\R'}(\r')|^2 U(|\r - \r'|) \, \d^2 r \, \d^2 r'. \label{eq:U_R_R}
\end{equation}
More details on the derivation are given in Appendix~\ref{sec:appendixB}\@. Fig.~\ref{fig:t_U}(b)--(c) shows the on-site interaction $U_0$ (orange circles) and nearest-neighbor interaction $U_1$ (orange squares) as function of twist angle for two different dielectric environments, i.e.\ freestanding and encapsulated in hexagonal boron nitride (hBN). The on-site interaction $U_0$ clearly dominates over the (next-) nearest-neighbor interactions. However, while hopping of IXs between lattice sites is quenched for small twist angles as expected due to increasing localization, we also find the on-site interaction $U_0$ to decrease with the twist angle, which might seem counterintuitive. This is because only the relative extent gets smaller for small twists, meaning stronger localization. At the same time the absolute spatial extent of the IX wave function, as given by $\langle |\r|\rangle_{w_{\R}}$ in Eq.~\eqref{eq:abs_extent} actually increases, thereby weakening the on-site interaction.

\begin{table*}
\caption{Parameters for the dielectric screening. Relative permittivities $\epsilon_\mathrm{r}$ taken from \cite{laturia_dielectric_2018} are given by the geometric mean of in-plane and out-of-plane dielectric constants. Values of the 2D polarizability $\chi_\mathrm{2D}$ are in units of nm.}
\begin{ruledtabular}
\begin{tabular}{lrrrrrrr}
Material								& MoS$_2$	& WS$_2$ &MoSe$_2$ 	& WSe$_2$ 		& hBN	& SiO$_2$	& air \\
\midrule
$\epsilon_\mathrm{r}$	& 	9.69\footnotemark[1] 			& 9.24\footnotemark[1] 		&11\footnotemark[1] 				& 	10.64 \footnotemark[1]			& 5.1\footnotemark[1] 	& 3.9 			& 1 \\
$\chi_\mathrm{2D}$ 		& 	0.711\footnotemark[2] 	 			& 	0.639\footnotemark[2] 	 		&0.846\footnotemark[2] 			& 	0.757\footnotemark[2] 				& 	-		& 	-				& -

\end{tabular}
\end{ruledtabular}
\footnotetext[1]{Ref.~\cite{laturia_dielectric_2018}}
\footnotetext[2]{Ref.~\cite{kylanpaa_binding_2015}}
\label{tab:screenings}
\end{table*}

\section{Dielectric screening} \label{sec:mod_screening}

\begin{figure}
    \centering
    \includegraphics[width=\linewidth]{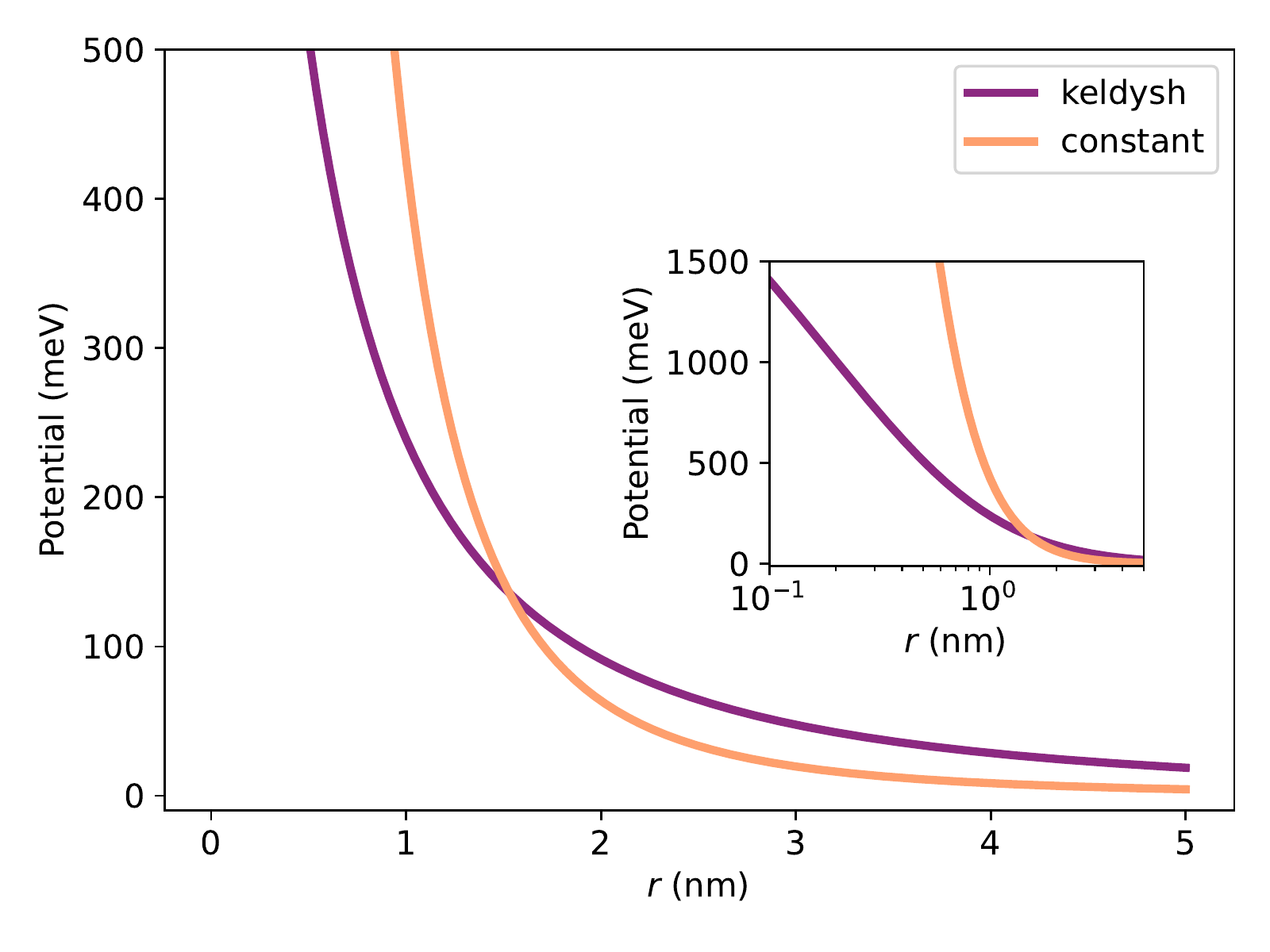}
    \caption{Dipolar potential as a function of IX distance $r$, comparing the effect of screening by a constant permittivity $\eps$ and non-local screening according to Eqs.~\eqref{eq:U_r} and \eqref{eq:U_r_mod}. For small distances the potential is reduced, changing from a $1/r$ dependence to a $\log(1/r)$ dependence (illustrated in the inset on a semilogarithmic scale). For larger distances the $1/r^3$ dependence remains.}
    \label{fig:mod_U_r}
\end{figure}

In Eq.~\eqref{eq:U_r} the dipolar interaction potential contains the dielectric environment of IXs solely as a constant relative permittivity $\epsilon_\mathrm{r}$ that is obtained by averaging the permittivities of the two constituent layers and super- and substrates \cite{wu_hubbard_2018,lagoin_key_2021}. However, a constant permittivity is better suited to capture effects in homogeneous media, while in heterostructures of atomically thin materials, the dielectric environment changes on small length scales, and in many situations, a more detailed modeling of screening effects is required \cite{steinke_noninvasive_2017, thygesen_calculating_2017, borghardt_engineering_2017, velicky_two-dimensional_2017, florian_dielectric_2018}. To improve on the constant-screening approach, we take into account non-local screening effects via a non-local Rytova-Keldysh potential. This approach was successfully used to model the Coulomb interaction of excitons, trions, and other complexes in TMD monolayers and heterostructures \cite{van_tuan_coulomb_2018,rodin_excitons_2014,kylanpaa_binding_2015,berman_superfluidity_2017,cudazzo_dielectric_2011,berkelbach_theory_2013,berghauser_analytical_2014}. We apply this description by modifying $U(r)$ in momentum space. The Fourier-transformed dipolar potential that determines the IX's interaction is given by
\begin{equation}
U(q) = \int_{\mathbb{R}^2} U(r)\e^{-\i \r \cdot \q}\, \d^2 r=\frac{e^2}{\epsilon_\mathrm{r}\epsilon_0}\frac{1}{q}(1-\e^{-dq}).\label{eq:U_FFT}
\end{equation}
Eq.~\eqref{eq:U_FFT} enables us to replace the constant permittivity $\epsilon_\mathrm{r} \epsilon_0$ by the non-local $\epsilon (q) = \epsilon_0 \kappa(1+r_0q)$, where $\kappa=(\epsilon_\mathrm{r}^\mathrm{t} + \epsilon_\mathrm{r}^\mathrm{b})/2$ is the mean relative permittivity of the dielectric environment around the heterostructure, $r_0=2\pi\chi_\mathrm{2D}/\kappa$ defines a length scale on which the potential is modified, and $\chi_\mathrm{2D}$ is the 2D polarizability of the heterostructure \cite{rodin_excitons_2014, kylanpaa_binding_2015}. We average the 2D polarizabilities of the two materials constituting the bilayer to find an effective description of the system. In real space, the modified dipolar potential becomes
\begin{equation}
  \tilde{U}(r) = \frac{e^2}{2\pi \kappa \epsilon_0}\int_0^\infty \frac{1-\e^{-dq}}{1+r_0q}J_0(rq)\,\d q \label{eq:U_r_mod}
\end{equation}
with the Bessel function of the first kind of order zero, $J_0$. The integral is evaluated numerically and the non-local dipolar potential is shown in Fig.~\ref{fig:mod_U_r} (purple). A comparison with the potential based on the constant screening (orange) reveals that the non-local screening results in lower dipolar potential strengths at small distances, changing the $1/r$ dependence to a logarithmic divergence (see inset in Fig.~\ref{fig:mod_U_r}) from which we expect reduced on-site interactions. For larger distances the modified potential retains the long-range $1/r^3$ dependence.

Figs.~\ref{fig:t_U}(b)--(c) (purple lines) show calculations of $U_n$ with the modified dipolar potential for two different dielectric environments. In general, the modified potential predicts smaller on-site interaction strengths. The amount of the reduction obtained from the non-local screening model depends on the actual dielectric environment as well as on the twist angle. For freestanding monolayers, the modified potential even predicts a stronger on-site repulsion than the local case, while for larger twist angles the effect of dielectric screening becomes more prominent. Furthermore, while the influence of on-site interaction $U_0$ clearly remains dominant, with the non-local screening model we find the strength of nearest-neighbor interactions $U_1$ strongly increased relative to on-site repulsion. This has implications for the occurrence of correlated IX phases as will be discussed in the following Section.

\section{Phases of interlayer excitons}
\label{sec:discussion}

Having established a connection between the moiré IX system and the generalized BH model, we now discuss the emergence of correlated states of IXs and possible signatures of these phases in experimental setups.
The on-site interaction $U_0$ and the nearest-neighbor hopping $t_1$ dominate the behavior, and we consider as a first approximation the standard BH Hamiltonian
\begin{equation}
    H_0 = -t_1 \sum_{\langle i, j \rangle} c_i^\dag c_j +  \frac{U_0}{2} \sum_{i} n_i (n_i-1), \label{eq:bh_0_order}
\end{equation}
where $\langle i, j \rangle$ denotes summation only over nearest neighbors, and $n_i$ is the Boson-number operator at site $i$. This Bose-Hubbard model in two dimensions is often treated in a mean-field approximation and predicts two distinct phases: For dominating on-site interactions ($t_1/U_0 \ll 1$)  the system is found to be in a Mott insulating phase at densities that correspond to commensurate fillings of the lattice with zero compressibility and suppressed particle movement \cite{fisher_boson_1989,wu_hubbard_2018,lagoin_key_2021}. If hopping dominates, a superfluid phase is found, which is a coherent IX state delocalized over the lattice. The superfluid phase is also found in the Mott regime at non-commensurate fillings, where an excess of particles moves freely over a commensurate, insulating background. A variety of other methods have been employed to treat the BH model beyond the mean-field level including Monte-Carlo \cite{capogrosso-sansone_monte_2008, bogner_variational_2019, kato_modification_2007} and path-integral methods \cite{polkovnikov_phase_2010} as well as matrix-product-state \cite{weiss_kibble-zurek_2018,iblisdir_matrix_2007} and density-matrix renormalization-group approaches \cite{garca-ripoll_variational_2004} in order to extract finite temperature properties of the system.

\begin{figure}[b]
    \centering
    \includegraphics[width=\linewidth]{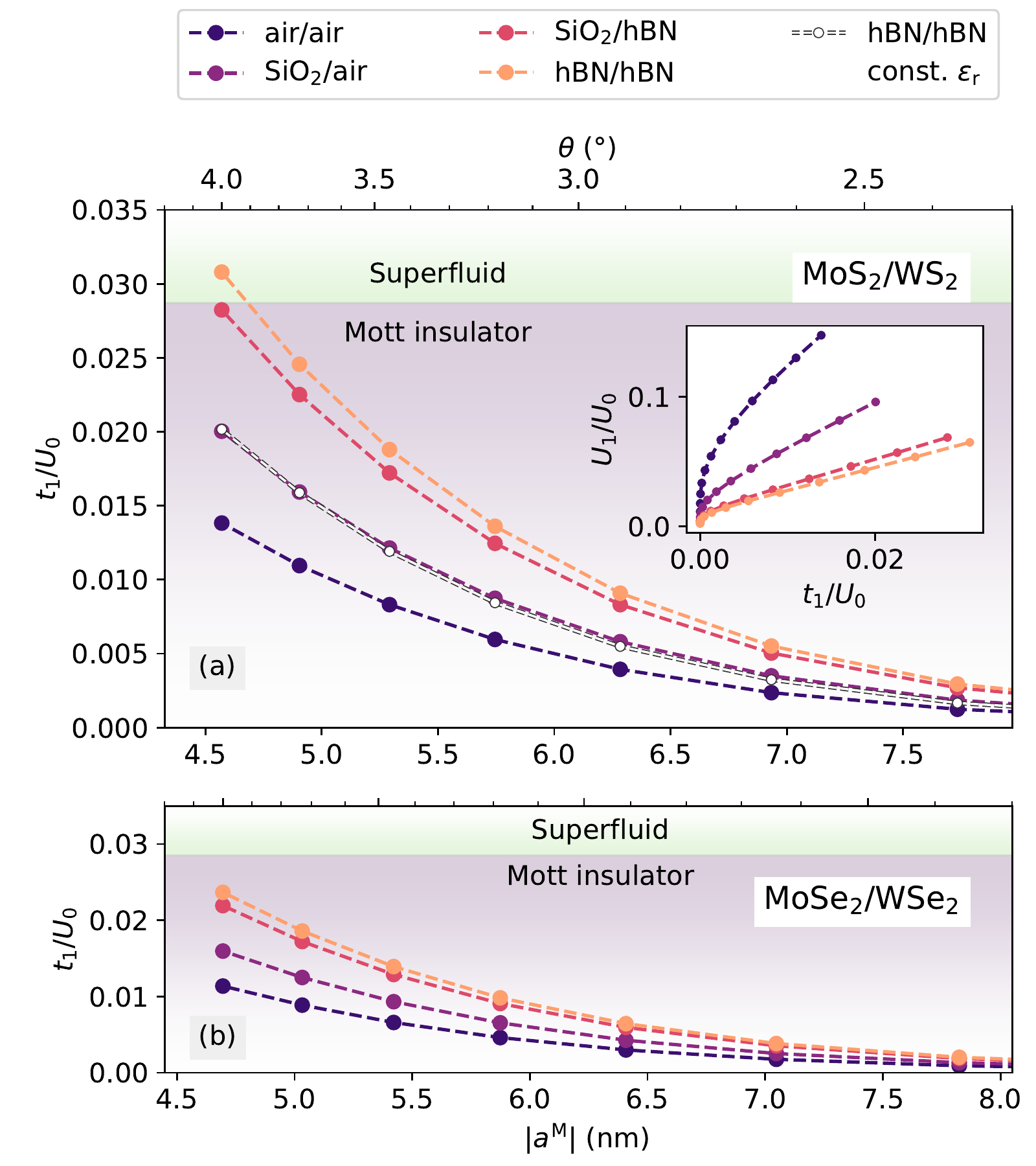}
    \caption{(a) Bose-Hubbard parameters $t_1/U_0$ for MoS$_2$/WS$_2$ as a function of twist angle and for different combinations of dielectric environment. The shaded areas indicate Mott and superfluid IX phases as predicted by a mean-field treatment of the BH system for a unit filling factor ($n=1$). Inset: Parametric plot of $U_1/U_0$ and $t_1/U_0$ as function of twist angle. The values for intermediate to large angles indicate the possibility of density-wave phases at fractional fillings. (b)  Parameters $t_1/U_0$ for MoSe$_2$/WeS$_2$ as a function of twist angle. In comparison to (a), for all dielectric environments the system remains in the Mott phase for all examined twist angles.}
    \label{fig:t_over_U}
\end{figure}

We first discuss results for the MoS$_2$/WS$_2$ material combination. In Fig.~\ref{fig:t_over_U}(a) we show the ratio $t_1/U_0$ obtained from our calculations as a function of twist angle for different commonly encountered combinations of sub- and superstrates. The shaded areas (violet/green) indicate the phase transition from the Mott insulating to the superfluid state for a filling factor of $n=1$ that we obtain from numerically solving the mean-field BH model. We would like to emphasize that most theoretical analysis treat the BH model in a grand-canonical ensemble description that allows for fluctuations in the particle number with an effective Hamiltonian $H_0 -\mu N$. Here, we consider situations, in which a fixed density of IXs is excited in the TMD heterobilayers that is allowed to equilibrate before radiative recombination, which corresponds to the experimental situation following optical excitation. In this spirit, we evaluate the mean-field model for a fixed filling factor of $n=1$, which corresponds to a canonical ensemble picture \cite{gygi_simulations_2006, ohgoe_ground-state_2012}. The results in Fig.~\ref{fig:t_over_U}(a) suggest that the MoS$_2$/WS$_2$ heterobilayer system will be deeply in the Mott regime for most considered twist angles and dielectric environments. The signature of this regime is a strongly reduced IX diffusion whenever excitation powers create IX densities that correspond to a filling of $n=1$, a change of system behavior that has only recently been observed experimentally in \cite{wang_diffusivity_2021} for a MoSe$_2$/WSe$_2$ system.
Numerical results for  MoSe$_2$/WSe$_2$ are shown in Fig.~\ref{fig:t_over_U}(b) and indicate that the selenide bilayer system remains in the Mott phase for all examined twist angles owing to a reduced mobility of the IXs caused by a larger lattice constant.

For larger twist angles and substantial dielectric screening, as realized by hBN encapsulation or capping on a silicon substrate, also the superfluid regime is accessible for the MoS$_2$/WS$_2$ heterostructure. Potentially, as the initially excited IX density decays due to radiative recombination, transient commensurate fillings can be revealed in the coherence of emitted light that allows to detect the presence or absence of an excess coherent superfluid phase on top of the commensurate, Mott-insulating background. As recently demonstrated, in more complex experimental configurations the presence of insulating IX phases can be detected by altered reflectivity and a blueshift of the IX photoluminescence due to changes in the dielectric screening in the heterobilayer system depending on the IX phase \cite{xu_correlated_2020, gu_dipolar_2022}. We point out that the trapping of IXs in the energetic minima of the moiré potential as discussed in Section \ref{sec:moire_IX} is distinct from the localization of IXs in the Mott phase, which relies on the mutual repulsion of IXs and can arise also for weak moiré potentials. 

For the hBN-encapsulated bilayer, we have added the local screening result as discussed in Section \ref{sec:moire_IX} as a gray curve to Fig.~\ref{fig:t_over_U}(a) for comparison. It becomes clear that the increased screening provided by the non-local potential has a significant impact on the phase boundaries. In our case, the associated underestimation of the dielectric screening would predict a Mott-phase even at the highest considered twist angles. Lagoin \emph{et al.}\ \cite{lagoin_key_2021} have discussed the possibility to insert layers of insulating hBN between the monolayers of the heterostructure to reduce the Coulomb interaction between IXs, making the superfluid regime more accessible. While it is an interesting avenue for further investigations, this approach also changes the strength of the moiré potential itself that is the prerequisite for a Bose-Hubbard description of the system.  We refrain from considering filling factors with more than one IX per moiré unit cell, as the role of many-particle effects, such as biexciton formation, may play an important role that is not captured in our current approach. Further details on the validity regime are discussed in Appendix~\ref{sec:density_estimates}\@.

Our approach to calculate the two-particle interaction beyond the on-site interaction matrix element $U_0$ allows drawing conclusions towards more exotic correlated bosonic states. Previous studies have reported density-wave phases for particular fractional fillings ($n= 1/3,\, 1/2, \, 2/3, \, \ldots$) in the presence of non-negligible $U_1$ \cite{regan_mott_2020, wilson_excitons_2021,ohgoe_ground-state_2012, miao_strong_2021,huang_correlated_2021}. In the inset of Fig.~\ref{fig:t_over_U} we show a parametric plot of $U_1/U_0$ over $t_1/U_0$ for varying twist angle, where values close to the origin correspond to smaller angles. We see, especially in the case of weaker dielectric screening, that the nearest-neighbor interaction can indeed become comparable to the on-site interaction $U_0$ and the hopping amplitude $t_1$, opening up the possibility for the formation of more complex phases beyond the superfluid and Mott-insulating phases. While this work is mainly directed at establishing a connection between the moiré physics of IXs and the Bose-Hubbard model, we do not delve further into the phase diagram at fractional fillings. We have, however, verified on a mean-field level that the results shown in Fig.~\ref{fig:t_over_U} remain valid even in the presence of next-nearest neighbor hopping. Our first results indicate that interlayer excitons in the tunable potential landscape of moiré TMD heterostructures provide a fascinating gateway to study phase transitions in flat-band systems \cite{huber_bose_2010, hui_superfluidity_2017}, and more complex correlated phases like density wave and supersolids. We are convinced that the underlying physics will be addressed both in theoretical and experimental work in the near future.

\section{Conclusion}
In conclusion, we have examined the moiré-Bose-Hubbard physics of IXs in MoS$_2$/WS$_2$ and MoSe$_2$/WSe$_2$ heterobilayers. A connection between the properties of IXs traversing the long-range potential landscape created by the moiré pattern and the Bose-Hubbard model is established by explicitly calculating the two-particle interaction and hopping parameters from a Wannier representation of the IX wave functions. Screening from the dielectric environment is treated within a Rytova-Keldysh-model approach, from which we find a significant reduction of the repulsive dipolar potential in comparison to the approach with constant permittivity that considers averages of dielectric constants of the different layers. Mean-field treatment of the Bose-Hubbard Hamiltonian predicts the existence of a Mott phase at unit fill factor and small twist angles due to dominating on-site interaction strength $U_0$. However, a transition to the superfluid state appears possible in the sulfur-based system for the relevant cases of hBN-encapsulated bilayers, and hBN capped bilayers on a silicon substrate, in which case the screening sufficiently lowers the dipolar interaction strength.

While our model approach offers first glimpses into the exciting correlated-state properties of excitons in van der Waals heterostructures, in which the twist angle can be used as a tuning knob for the Bose-Hubbard parameters, it also reveals limitations that are inherent to current models, such as multi-band and biexcitonic effects that will likely alter the physics at larger filling factors. A particularly promising platform to further explore correlated phases in moiré-Bose Hubbard systems are cavity-embedded bilayers, in which IX and photons form IX polaritons with even greater control over the ratio between dipolar interaction and hopping. Furthermore, it will be interesting to study effects of local reconstructions, which have been shown to strongly impact the moiré physics at low twists angles \cite{zhao_excitons_2022}.

\section*{Acknowledgements}
We would like to thank Alexander Steinhoff for many useful discussions. Funding is acknowledged from the Deutsche Forschungsgemeinschaft (German Research Foundation) via the priority program SP2244 (project Gi-1121/4-1) and the graduate school 2247. F.L.\ further acknowledges funding by the central research development fund (CRDF) of the University of Bremen.

\appendix
\section{Estimates of IX Densities}
\label{sec:density_estimates}
We provide estimates under which conditions the assumption that the IXs can be treated as composite particles is valid. For Bosons not confined to individual lattice sites, we note that particles with average thermal energy $E = \kb T$ can approach each other due to their mutual dipolar repulsion up to a distance $r_E$ that is determined by $U(r_E) = E$. We aim at densities and temperatures such that IXs only see each other's dipole field ($\sim 1/r^3$) and not their inner structure. This is satisfied if we require $d < r_E$ from which we find (with $c = e^2/4\pi\eps$)
\begin{equation}
  U(r_E) = c \left(\frac{2}{r_E} - \frac{2}{\sqrt{d^2 + r_E^2}}\right)\approx \frac{c d^2}{r_E^3}.\label{eq:U_r_estimate}
\end{equation}
From $U(r_E) = \kb T$ follows
\begin{eqnarray}
&\dfrac{\kb T d}{c} = \dfrac{d^3}{r_E^3}  \ll 1\nonumber\\
\Longrightarrow \qquad &\kb T \ll \dfrac{e^2}{4\pi \eps d}.
\end{eqnarray}
With exemplary values $d = 0.647\,\mathrm{nm}$ and $\er = 10$ this results in a temperature limit of $T \ll 2500\,\mathrm{K}$, which is met for all experimental conditions.

\begin{figure}[h]
    \centering
    \includegraphics[width=\linewidth]{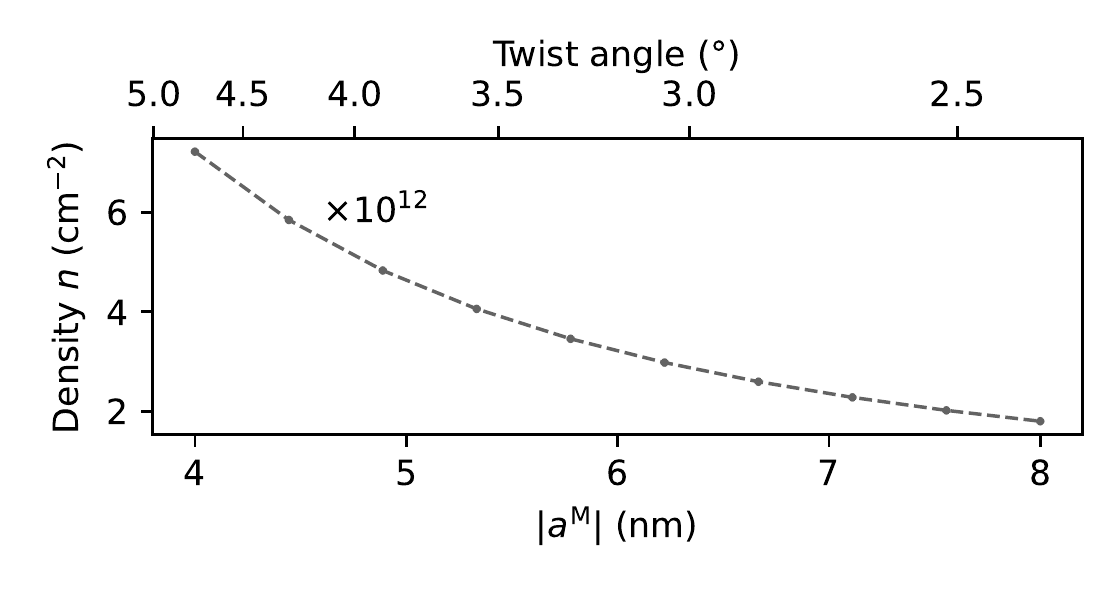}
    \caption{Maximal density as function of moiré lattice constant to have maximal one IX per potential minimum.}
    \label{fig:density_estimate}
\end{figure}

In the case of IXs confined to the potential minima in each moiré unit cell, Fig.~\ref{fig:density_estimate} shows an upper density limit that corresponds to a maximum number of one IX per unit cell as a function of the moiré lattice constant $|\aM|$. The estimate can easily be determined from the unit cell geometry and is given by $n \leq \frac{2}{\sqrt{3}|\aM|}$.

\section{Calculation of the Hubbard parameters}
\label{sec:appendixB}

The Hubbard parameters are the coefficients of the second quantization operators derived from $H_0$ and $H_1$.
For the one-particle operator $H_0$, these coefficients are
\begin{equation}
    t_{\R, \R'} = \braket{\R| H_0 | \R'}.
\end{equation}
Using the localized Wannier functions from Eq.~\eqref{eq:wannier_def} as the basis, this expression becomes
\begin{align}
    t_{\R, \R'} &= \int_{\mathbb{R}^2} w^*_{\R}(\r) H_0 w_{\R'}(\r)\d^2 r\nonumber\\
                &= \frac{1}{N}\sum_{\Q, \Q'} \int_{\mathbb{R}^2} \e^{\i \Q\R}\chi^*_\Q(\r)H_0\e^{-\i\Q'\R'}\chi_{\Q'}(\r)\d^2 r.
\end{align}
As the Bloch functions $\chi_\Q(\r)$ by definition are eigenfunctions of $H_0$ with eigenvalues $E_\Q$, we obtain
\begin{align}
    t_{\R, \R'} &= \frac{1}{N} \sum_{\Q, \Q'}
    \e^{\i(\Q\R - \Q'\R')} E_{\Q'}
        \int_{\mathbb{R}^2}
            \chi^*_\Q(\r)
            \chi_{\Q'}(\r)
        \d^2 r\nonumber\\
    &=
    \frac{1}{N} \sum_{\Q, \Q'}
    \e^{\i(\Q\R - \Q'\R')} E_{\Q'} \delta_{\Q, \Q'}\nonumber\\
    &=
    \frac{1}{N} \sum_{\Q}
    \e^{\i(\R - \R')\Q} E_{\Q} =: t(\R - \R'),
\end{align}
which is the result used in Eq.~\eqref{eq:tn}. The coefficients of the two-particle operator $H_1$ are
\begin{equation}
    U_{\R_1, \R_2, \R_3, \R_4} = \bra{\R_1}^{(i)}\bra{\R_2}^{(j)} H_1 \ket{\R_3}^{(j)} \ket{\R_4}^{(i)}.
\end{equation}
With the Wannier functions as the basis and the vectors $\r$ and $\r'$ being the positions of particles $(i)$ and $(j)$, respectively, the coefficients become
\begin{multline}
    U_{\R_1, \R_2, \R_3, \R_4} = \iint_{\mathbb{R}^2} w^*_{\R_1}(\r)w^*_{\R_2}(\r') U(|\r - \r'|) \\w_{\R_3}(\r')w_{\R_4}(\r)\, \d^2 r\, \d^2 r'.
\end{multline}
Due to the strong localization of the Wannier functions, the product $w_{\R}(\r)w_{\R'}(\r)$ is non-negligible only if $\R = \R'$.
This justifies the approximation $\R_1 = \R_4 := \R,~\R_2 = \R_3 =: \R'$ and results in the two-particle coefficient
\begin{align}
    U_{\R, \R'} &= \iint_{\mathbb{R}^2} w^*_{\R}(\r)w^*_{\R'}(\r') U(|\r - \r'|) \nonumber\\
                &\qquad\qquad\qquad w_{\R'}(\r')w_{\R}(\r)\, \d^2 r\, \d^2 r'\nonumber\\
                &=\iint_{\mathbb{R}^2} |w_{\R}(\r)|^2 |w_{\R'}(\r')|^2 U(|\r - \r'|) \, \d^2 r\, \d^2 r', \label{eq:U_app}
\end{align}
which is the result used in Eq.~\eqref{eq:U_R_R}.
Finally, we note that care must be taken during the numerical evaluation of Eq.~\eqref{eq:U_app} as the term $U(|\r - \r'|)$ features singularities. However, as the  integration area is the two-dimensional plane, and the singularities are of the type $1/r$ the singularities can be lifted via a transformation to polar coordinates. For that it is useful to first express $\r$ in Eq.~\eqref{eq:U_app} in terms of a new variable $\tr = \r - \r'$. By furthermore shifting the Wannier functions so that $\R'$ lies in the origin and the distance between the functions' centers remains the same, the following expression for $U$ is obtained:
\begin{align}
    U_{\R, \R'} &= \iint_{\mathbb{R}^2} |w_{\R - \R'}(\tr + \r')|^2 |w_{0}(\r')|^2 U(\tilde{r}) \, \d^2 \tilde{r} \, \d^2 r'\nonumber\\
                &=: U(\R - \R').
\end{align}
Finally, the transformation to polar coordinates can be applied, yielding
\begin{multline}
U(\R - \R') = \frac{e^2}{4\pi\epsilon\ind{r}\epsilon_0}\iint_{\mathbb{R}^2} |w_{\R - \R'}(\tr + \r')|^2 |w_{0}(\r')|^2 \\\left(r' - \frac{r'\tilde{r}}{\sqrt{\tilde{r}^2 + d^2}}\right) \, \d \tilde{r}\,\d \tilde{\varphi} \, \d r'\,\d\varphi',
\end{multline}
which is free of singularities and can be readily evaluated.

\bibliography{Moire_Bose_Hubbard}

\begin{thebibliography}{63}%
\makeatletter
\providecommand \@ifxundefined [1]{%
 \@ifx{#1\undefined}
}%
\providecommand \@ifnum [1]{%
 \ifnum #1\expandafter \@firstoftwo
 \else \expandafter \@secondoftwo
 \fi
}%
\providecommand \@ifx [1]{%
 \ifx #1\expandafter \@firstoftwo
 \else \expandafter \@secondoftwo
 \fi
}%
\providecommand \natexlab [1]{#1}%
\providecommand \enquote  [1]{``#1''}%
\providecommand \bibnamefont  [1]{#1}%
\providecommand \bibfnamefont [1]{#1}%
\providecommand \citenamefont [1]{#1}%
\providecommand \href@noop [0]{\@secondoftwo}%
\providecommand \href [0]{\begingroup \@sanitize@url \@href}%
\providecommand \@href[1]{\@@startlink{#1}\@@href}%
\providecommand \@@href[1]{\endgroup#1\@@endlink}%
\providecommand \@sanitize@url [0]{\catcode `\\12\catcode `\$12\catcode
  `\&12\catcode `\#12\catcode `\^12\catcode `\_12\catcode `\%12\relax}%
\providecommand \@@startlink[1]{}%
\providecommand \@@endlink[0]{}%
\providecommand \url  [0]{\begingroup\@sanitize@url \@url }%
\providecommand \@url [1]{\endgroup\@href {#1}{\urlprefix }}%
\providecommand \urlprefix  [0]{URL }%
\providecommand \Eprint [0]{\href }%
\providecommand \doibase [0]{https://doi.org/}%
\providecommand \selectlanguage [0]{\@gobble}%
\providecommand \bibinfo  [0]{\@secondoftwo}%
\providecommand \bibfield  [0]{\@secondoftwo}%
\providecommand \translation [1]{[#1]}%
\providecommand \BibitemOpen [0]{}%
\providecommand \bibitemStop [0]{}%
\providecommand \bibitemNoStop [0]{.\EOS\space}%
\providecommand \EOS [0]{\spacefactor3000\relax}%
\providecommand \BibitemShut  [1]{\csname bibitem#1\endcsname}%
\let\auto@bib@innerbib\@empty
%</preamble>
\bibitem [{\citenamefont {Wilson}\ \emph {et~al.}(2021)\citenamefont {Wilson},
  \citenamefont {Yao}, \citenamefont {Shan},\ and\ \citenamefont
  {Xu}}]{wilson_excitons_2021}%
  \BibitemOpen
  \bibfield  {author} {\bibinfo {author} {\bibfnamefont {N.~P.}\ \bibnamefont
  {Wilson}}, \bibinfo {author} {\bibfnamefont {W.}~\bibnamefont {Yao}},
  \bibinfo {author} {\bibfnamefont {J.}~\bibnamefont {Shan}},\ and\ \bibinfo
  {author} {\bibfnamefont {X.}~\bibnamefont {Xu}},\ }\bibfield  {title}
  {\bibinfo {title} {Excitons and emergent quantum phenomena in stacked {2D}
  semiconductors},\ }\href {https://doi.org/10.1038/s41586-021-03979-1}
  {\bibfield  {journal} {\bibinfo  {journal} {Nature}\ }\textbf {\bibinfo
  {volume} {599}},\ \bibinfo {pages} {383} (\bibinfo {year}
  {2021})}\BibitemShut {NoStop}%
\bibitem [{\citenamefont {Kennes}\ \emph {et~al.}(2021)\citenamefont {Kennes},
  \citenamefont {Claassen}, \citenamefont {Xian}, \citenamefont {Georges},
  \citenamefont {Millis}, \citenamefont {Hone}, \citenamefont {Dean},
  \citenamefont {Basov}, \citenamefont {Pasupathy},\ and\ \citenamefont
  {Rubio}}]{kennes_moire_2021}%
  \BibitemOpen
  \bibfield  {author} {\bibinfo {author} {\bibfnamefont {D.~M.}\ \bibnamefont
  {Kennes}}, \bibinfo {author} {\bibfnamefont {M.}~\bibnamefont {Claassen}},
  \bibinfo {author} {\bibfnamefont {L.}~\bibnamefont {Xian}}, \bibinfo {author}
  {\bibfnamefont {A.}~\bibnamefont {Georges}}, \bibinfo {author} {\bibfnamefont
  {A.~J.}\ \bibnamefont {Millis}}, \bibinfo {author} {\bibfnamefont
  {J.}~\bibnamefont {Hone}}, \bibinfo {author} {\bibfnamefont {C.~R.}\
  \bibnamefont {Dean}}, \bibinfo {author} {\bibfnamefont {D.~N.}\ \bibnamefont
  {Basov}}, \bibinfo {author} {\bibfnamefont {A.~N.}\ \bibnamefont
  {Pasupathy}},\ and\ \bibinfo {author} {\bibfnamefont {A.}~\bibnamefont
  {Rubio}},\ }\bibfield  {title} {\bibinfo {title} {Moiré heterostructures as
  a condensed-matter quantum simulator},\ }\href
  {https://doi.org/10.1038/s41567-020-01154-3} {\bibfield  {journal} {\bibinfo
  {journal} {Nature Physics}\ }\textbf {\bibinfo {volume} {17}},\ \bibinfo
  {pages} {155} (\bibinfo {year} {2021})}\BibitemShut {NoStop}%
\bibitem [{\citenamefont {Wu}\ \emph {et~al.}(2017)\citenamefont {Wu},
  \citenamefont {Lovorn},\ and\ \citenamefont
  {MacDonald}}]{wu_topological_2017}%
  \BibitemOpen
  \bibfield  {author} {\bibinfo {author} {\bibfnamefont {F.}~\bibnamefont
  {Wu}}, \bibinfo {author} {\bibfnamefont {T.}~\bibnamefont {Lovorn}},\ and\
  \bibinfo {author} {\bibfnamefont {A.}~\bibnamefont {MacDonald}},\ }\bibfield
  {title} {\bibinfo {title} {Topological exciton bands in moiré
  heterojunctions},\ }\href {https://doi.org/10.1103/PhysRevLett.118.147401}
  {\bibfield  {journal} {\bibinfo  {journal} {Physical Review Letters}\
  }\textbf {\bibinfo {volume} {118}},\ \bibinfo {pages} {147401} (\bibinfo
  {year} {2017})}\BibitemShut {NoStop}%
\bibitem [{\citenamefont {Wu}\ \emph {et~al.}(2018{\natexlab{a}})\citenamefont
  {Wu}, \citenamefont {Lovorn},\ and\ \citenamefont
  {MacDonald}}]{wu_theory_2018}%
  \BibitemOpen
  \bibfield  {author} {\bibinfo {author} {\bibfnamefont {F.}~\bibnamefont
  {Wu}}, \bibinfo {author} {\bibfnamefont {T.}~\bibnamefont {Lovorn}},\ and\
  \bibinfo {author} {\bibfnamefont {A.~H.}\ \bibnamefont {MacDonald}},\
  }\bibfield  {title} {\bibinfo {title} {Theory of optical absorption by
  interlayer excitons in transition metal dichalcogenide heterobilayers},\
  }\href {https://doi.org/10.1103/PhysRevB.97.035306} {\bibfield  {journal}
  {\bibinfo  {journal} {Physical Review B}\ }\textbf {\bibinfo {volume} {97}},\
  \bibinfo {pages} {035306} (\bibinfo {year} {2018}{\natexlab{a}})}\BibitemShut
  {NoStop}%
\bibitem [{\citenamefont {Tran}\ \emph {et~al.}(2019)\citenamefont {Tran},
  \citenamefont {Moody}, \citenamefont {Wu}, \citenamefont {Lu}, \citenamefont
  {Choi}, \citenamefont {Kim}, \citenamefont {Rai}, \citenamefont {Sanchez},
  \citenamefont {Quan}, \citenamefont {Singh}, \citenamefont {Embley},
  \citenamefont {Zepeda}, \citenamefont {Campbell}, \citenamefont {Autry},
  \citenamefont {Taniguchi}, \citenamefont {Watanabe}, \citenamefont {Lu},
  \citenamefont {Banerjee}, \citenamefont {Silverman}, \citenamefont {Kim},
  \citenamefont {Tutuc}, \citenamefont {Yang}, \citenamefont {MacDonald},\ and\
  \citenamefont {Li}}]{tran_evidence_2019}%
  \BibitemOpen
  \bibfield  {author} {\bibinfo {author} {\bibfnamefont {K.}~\bibnamefont
  {Tran}}, \bibinfo {author} {\bibfnamefont {G.}~\bibnamefont {Moody}},
  \bibinfo {author} {\bibfnamefont {F.}~\bibnamefont {Wu}}, \bibinfo {author}
  {\bibfnamefont {X.}~\bibnamefont {Lu}}, \bibinfo {author} {\bibfnamefont
  {J.}~\bibnamefont {Choi}}, \bibinfo {author} {\bibfnamefont {K.}~\bibnamefont
  {Kim}}, \bibinfo {author} {\bibfnamefont {A.}~\bibnamefont {Rai}}, \bibinfo
  {author} {\bibfnamefont {D.~A.}\ \bibnamefont {Sanchez}}, \bibinfo {author}
  {\bibfnamefont {J.}~\bibnamefont {Quan}}, \bibinfo {author} {\bibfnamefont
  {A.}~\bibnamefont {Singh}}, \bibinfo {author} {\bibfnamefont
  {J.}~\bibnamefont {Embley}}, \bibinfo {author} {\bibfnamefont
  {A.}~\bibnamefont {Zepeda}}, \bibinfo {author} {\bibfnamefont
  {M.}~\bibnamefont {Campbell}}, \bibinfo {author} {\bibfnamefont
  {T.}~\bibnamefont {Autry}}, \bibinfo {author} {\bibfnamefont
  {T.}~\bibnamefont {Taniguchi}}, \bibinfo {author} {\bibfnamefont
  {K.}~\bibnamefont {Watanabe}}, \bibinfo {author} {\bibfnamefont
  {N.}~\bibnamefont {Lu}}, \bibinfo {author} {\bibfnamefont {S.~K.}\
  \bibnamefont {Banerjee}}, \bibinfo {author} {\bibfnamefont {K.~L.}\
  \bibnamefont {Silverman}}, \bibinfo {author} {\bibfnamefont {S.}~\bibnamefont
  {Kim}}, \bibinfo {author} {\bibfnamefont {E.}~\bibnamefont {Tutuc}}, \bibinfo
  {author} {\bibfnamefont {L.}~\bibnamefont {Yang}}, \bibinfo {author}
  {\bibfnamefont {A.~H.}\ \bibnamefont {MacDonald}},\ and\ \bibinfo {author}
  {\bibfnamefont {X.}~\bibnamefont {Li}},\ }\bibfield  {title} {\bibinfo
  {title} {Evidence for moiré excitons in van der {Waals} heterostructures},\
  }\href {https://doi.org/10.1038/s41586-019-0975-z} {\bibfield  {journal}
  {\bibinfo  {journal} {Nature}\ }\textbf {\bibinfo {volume} {567}},\ \bibinfo
  {pages} {71} (\bibinfo {year} {2019})}\BibitemShut {NoStop}%
\bibitem [{\citenamefont {Brem}\ \emph {et~al.}(2020)\citenamefont {Brem},
  \citenamefont {Linderälv}, \citenamefont {Erhart},\ and\ \citenamefont
  {Malic}}]{brem_tunable_2020}%
  \BibitemOpen
  \bibfield  {author} {\bibinfo {author} {\bibfnamefont {S.}~\bibnamefont
  {Brem}}, \bibinfo {author} {\bibfnamefont {C.}~\bibnamefont {Linderälv}},
  \bibinfo {author} {\bibfnamefont {P.}~\bibnamefont {Erhart}},\ and\ \bibinfo
  {author} {\bibfnamefont {E.}~\bibnamefont {Malic}},\ }\bibfield  {title}
  {\bibinfo {title} {Tunable phases of moiré excitons in van der {Waals}
  heterostructures},\ }\href {https://doi.org/10.1021/acs.nanolett.0c03019}
  {\bibfield  {journal} {\bibinfo  {journal} {Nano Letters}\ }\textbf {\bibinfo
  {volume} {20}},\ \bibinfo {pages} {8534} (\bibinfo {year}
  {2020})}\BibitemShut {NoStop}%
\bibitem [{\citenamefont {Hubbard}\ and\ \citenamefont
  {Flowers}(1963)}]{hubbard_electron_1963}%
  \BibitemOpen
  \bibfield  {author} {\bibinfo {author} {\bibfnamefont {J.}~\bibnamefont
  {Hubbard}}\ and\ \bibinfo {author} {\bibfnamefont {B.~H.}\ \bibnamefont
  {Flowers}},\ }\bibfield  {title} {\bibinfo {title} {Electron correlations in
  narrow energy bands},\ }\href {https://doi.org/10.1098/rspa.1963.0204}
  {\bibfield  {journal} {\bibinfo  {journal} {Proc. R. Soc. Lond. A}\ }\textbf
  {\bibinfo {volume} {276}},\ \bibinfo {pages} {238} (\bibinfo {year}
  {1963})}\BibitemShut {NoStop}%
\bibitem [{\citenamefont {Fisher}\ \emph {et~al.}(1989)\citenamefont {Fisher},
  \citenamefont {Weichman}, \citenamefont {Grinstein},\ and\ \citenamefont
  {Fisher}}]{fisher_boson_1989}%
  \BibitemOpen
  \bibfield  {author} {\bibinfo {author} {\bibfnamefont {M.~P.~A.}\
  \bibnamefont {Fisher}}, \bibinfo {author} {\bibfnamefont {P.~B.}\
  \bibnamefont {Weichman}}, \bibinfo {author} {\bibfnamefont {G.}~\bibnamefont
  {Grinstein}},\ and\ \bibinfo {author} {\bibfnamefont {D.~S.}\ \bibnamefont
  {Fisher}},\ }\bibfield  {title} {\bibinfo {title} {Boson localization and the
  superfluid-insulator transition},\ }\href
  {https://doi.org/10.1103/PhysRevB.40.546} {\bibfield  {journal} {\bibinfo
  {journal} {Physical Review B}\ }\textbf {\bibinfo {volume} {40}},\ \bibinfo
  {pages} {546} (\bibinfo {year} {1989})}\BibitemShut {NoStop}%
\bibitem [{\citenamefont {Tang}\ \emph {et~al.}(2020)\citenamefont {Tang},
  \citenamefont {Li}, \citenamefont {Li}, \citenamefont {Xu}, \citenamefont
  {Liu}, \citenamefont {Barmak}, \citenamefont {Watanabe}, \citenamefont
  {Taniguchi}, \citenamefont {MacDonald}, \citenamefont {Shan},\ and\
  \citenamefont {Mak}}]{tang_simulation_2020}%
  \BibitemOpen
  \bibfield  {author} {\bibinfo {author} {\bibfnamefont {Y.}~\bibnamefont
  {Tang}}, \bibinfo {author} {\bibfnamefont {L.}~\bibnamefont {Li}}, \bibinfo
  {author} {\bibfnamefont {T.}~\bibnamefont {Li}}, \bibinfo {author}
  {\bibfnamefont {Y.}~\bibnamefont {Xu}}, \bibinfo {author} {\bibfnamefont
  {S.}~\bibnamefont {Liu}}, \bibinfo {author} {\bibfnamefont {K.}~\bibnamefont
  {Barmak}}, \bibinfo {author} {\bibfnamefont {K.}~\bibnamefont {Watanabe}},
  \bibinfo {author} {\bibfnamefont {T.}~\bibnamefont {Taniguchi}}, \bibinfo
  {author} {\bibfnamefont {A.~H.}\ \bibnamefont {MacDonald}}, \bibinfo {author}
  {\bibfnamefont {J.}~\bibnamefont {Shan}},\ and\ \bibinfo {author}
  {\bibfnamefont {K.~F.}\ \bibnamefont {Mak}},\ }\bibfield  {title} {\bibinfo
  {title} {Simulation of {Hubbard} model physics in {WSe$_2$}/{WS$_2$} moiré
  superlattices},\ }\href {https://doi.org/10.1038/s41586-020-2085-3}
  {\bibfield  {journal} {\bibinfo  {journal} {Nature}\ }\textbf {\bibinfo
  {volume} {579}},\ \bibinfo {pages} {353} (\bibinfo {year}
  {2020})}\BibitemShut {NoStop}%
\bibitem [{\citenamefont {Huang}\ \emph {et~al.}(2021)\citenamefont {Huang},
  \citenamefont {Wang}, \citenamefont {Miao}, \citenamefont {Wang},
  \citenamefont {Li}, \citenamefont {Lian}, \citenamefont {Taniguchi},
  \citenamefont {Watanabe}, \citenamefont {Okamoto}, \citenamefont {Xiao},
  \citenamefont {Shi},\ and\ \citenamefont {Cui}}]{huang_correlated_2021}%
  \BibitemOpen
  \bibfield  {author} {\bibinfo {author} {\bibfnamefont {X.}~\bibnamefont
  {Huang}}, \bibinfo {author} {\bibfnamefont {T.}~\bibnamefont {Wang}},
  \bibinfo {author} {\bibfnamefont {S.}~\bibnamefont {Miao}}, \bibinfo {author}
  {\bibfnamefont {C.}~\bibnamefont {Wang}}, \bibinfo {author} {\bibfnamefont
  {Z.}~\bibnamefont {Li}}, \bibinfo {author} {\bibfnamefont {Z.}~\bibnamefont
  {Lian}}, \bibinfo {author} {\bibfnamefont {T.}~\bibnamefont {Taniguchi}},
  \bibinfo {author} {\bibfnamefont {K.}~\bibnamefont {Watanabe}}, \bibinfo
  {author} {\bibfnamefont {S.}~\bibnamefont {Okamoto}}, \bibinfo {author}
  {\bibfnamefont {D.}~\bibnamefont {Xiao}}, \bibinfo {author} {\bibfnamefont
  {S.-F.}\ \bibnamefont {Shi}},\ and\ \bibinfo {author} {\bibfnamefont {Y.-T.}\
  \bibnamefont {Cui}},\ }\bibfield  {title} {\bibinfo {title} {Correlated
  insulating states at fractional fillings of the {WS$_2$}/{WSe$_2$} moiré
  lattice},\ }\href {https://doi.org/10.1038/s41567-021-01171-w} {\bibfield
  {journal} {\bibinfo  {journal} {Nature Physics}\ }\textbf {\bibinfo {volume}
  {17}},\ \bibinfo {pages} {715} (\bibinfo {year} {2021})}\BibitemShut
  {NoStop}%
\bibitem [{\citenamefont {Wu}\ \emph {et~al.}(2018{\natexlab{b}})\citenamefont
  {Wu}, \citenamefont {Lovorn}, \citenamefont {Tutuc},\ and\ \citenamefont
  {MacDonald}}]{wu_hubbard_2018}%
  \BibitemOpen
  \bibfield  {author} {\bibinfo {author} {\bibfnamefont {F.}~\bibnamefont
  {Wu}}, \bibinfo {author} {\bibfnamefont {T.}~\bibnamefont {Lovorn}}, \bibinfo
  {author} {\bibfnamefont {E.}~\bibnamefont {Tutuc}},\ and\ \bibinfo {author}
  {\bibfnamefont {A.~H.}\ \bibnamefont {MacDonald}},\ }\bibfield  {title}
  {\bibinfo {title} {Hubbard model physics in transition metal dichalcogenide
  moiré bands},\ }\href {https://doi.org/10.1103/PhysRevLett.121.026402}
  {\bibfield  {journal} {\bibinfo  {journal} {Physical Review Letters}\
  }\textbf {\bibinfo {volume} {121}},\ \bibinfo {pages} {026402} (\bibinfo
  {year} {2018}{\natexlab{b}})}\BibitemShut {NoStop}%
\bibitem [{\citenamefont {Slagle}\ and\ \citenamefont
  {Fu}(2020)}]{slagle_charge_2020}%
  \BibitemOpen
  \bibfield  {author} {\bibinfo {author} {\bibfnamefont {K.}~\bibnamefont
  {Slagle}}\ and\ \bibinfo {author} {\bibfnamefont {L.}~\bibnamefont {Fu}},\
  }\bibfield  {title} {\bibinfo {title} {Charge transfer excitations, pair
  density waves, and superconductivity in moiré materials},\ }\href
  {https://doi.org/10.1103/PhysRevB.102.235423} {\bibfield  {journal} {\bibinfo
   {journal} {Physical Review B}\ }\textbf {\bibinfo {volume} {102}},\ \bibinfo
  {pages} {235423} (\bibinfo {year} {2020})}\BibitemShut {NoStop}%
\bibitem [{\citenamefont {Pan}\ \emph {et~al.}(2020{\natexlab{a}})\citenamefont
  {Pan}, \citenamefont {Wu},\ and\ \citenamefont {Das~Sarma}}]{pan_band_2020}%
  \BibitemOpen
  \bibfield  {author} {\bibinfo {author} {\bibfnamefont {H.}~\bibnamefont
  {Pan}}, \bibinfo {author} {\bibfnamefont {F.}~\bibnamefont {Wu}},\ and\
  \bibinfo {author} {\bibfnamefont {S.}~\bibnamefont {Das~Sarma}},\ }\bibfield
  {title} {\bibinfo {title} {Band topology, {Hubbard} model, {Heisenberg}
  model, and {Dzyaloshinskii}-{Moriya} interaction in twisted bilayer
  {WSe$_2$}},\ }\href {https://doi.org/10.1103/PhysRevResearch.2.033087}
  {\bibfield  {journal} {\bibinfo  {journal} {Physical Review Research}\
  }\textbf {\bibinfo {volume} {2}},\ \bibinfo {pages} {033087} (\bibinfo {year}
  {2020}{\natexlab{a}})}\BibitemShut {NoStop}%
\bibitem [{\citenamefont {Pan}\ \emph {et~al.}(2020{\natexlab{b}})\citenamefont
  {Pan}, \citenamefont {Wu},\ and\ \citenamefont
  {Das~Sarma}}]{pan_quantum_2020}%
  \BibitemOpen
  \bibfield  {author} {\bibinfo {author} {\bibfnamefont {H.}~\bibnamefont
  {Pan}}, \bibinfo {author} {\bibfnamefont {F.}~\bibnamefont {Wu}},\ and\
  \bibinfo {author} {\bibfnamefont {S.}~\bibnamefont {Das~Sarma}},\ }\bibfield
  {title} {\bibinfo {title} {Quantum phase diagram of a moiré-{Hubbard}
  model},\ }\href {https://doi.org/10.1103/PhysRevB.102.201104} {\bibfield
  {journal} {\bibinfo  {journal} {Physical Review B}\ }\textbf {\bibinfo
  {volume} {102}},\ \bibinfo {pages} {201104} (\bibinfo {year}
  {2020}{\natexlab{b}})}\BibitemShut {NoStop}%
\bibitem [{\citenamefont {Torun}\ \emph {et~al.}(2018)\citenamefont {Torun},
  \citenamefont {Miranda}, \citenamefont {Molina-Sánchez},\ and\ \citenamefont
  {Wirtz}}]{torun_interlayer_2018}%
  \BibitemOpen
  \bibfield  {author} {\bibinfo {author} {\bibfnamefont {E.}~\bibnamefont
  {Torun}}, \bibinfo {author} {\bibfnamefont {H.~P.~C.}\ \bibnamefont
  {Miranda}}, \bibinfo {author} {\bibfnamefont {A.}~\bibnamefont
  {Molina-Sánchez}},\ and\ \bibinfo {author} {\bibfnamefont {L.}~\bibnamefont
  {Wirtz}},\ }\bibfield  {title} {\bibinfo {title} {Interlayer and intralayer
  excitons in {MoS$_2$}/{WS$_2$} and {MoSe$_2$}/{WSe$_2$} heterobilayers},\
  }\href {https://doi.org/10.1103/PhysRevB.97.245427} {\bibfield  {journal}
  {\bibinfo  {journal} {Physical Review B}\ }\textbf {\bibinfo {volume} {97}},\
  \bibinfo {pages} {245427} (\bibinfo {year} {2018})}\BibitemShut {NoStop}%
\bibitem [{\citenamefont {Jiang}\ \emph {et~al.}(2021)\citenamefont {Jiang},
  \citenamefont {Chen}, \citenamefont {Zheng}, \citenamefont {Zheng},\ and\
  \citenamefont {Pan}}]{jiang_interlayer_2021}%
  \BibitemOpen
  \bibfield  {author} {\bibinfo {author} {\bibfnamefont {Y.}~\bibnamefont
  {Jiang}}, \bibinfo {author} {\bibfnamefont {S.}~\bibnamefont {Chen}},
  \bibinfo {author} {\bibfnamefont {W.}~\bibnamefont {Zheng}}, \bibinfo
  {author} {\bibfnamefont {B.}~\bibnamefont {Zheng}},\ and\ \bibinfo {author}
  {\bibfnamefont {A.}~\bibnamefont {Pan}},\ }\bibfield  {title} {\bibinfo
  {title} {Interlayer exciton formation, relaxation, and transport in {TMD} van
  der {Waals} heterostructures},\ }\href
  {https://doi.org/10.1038/s41377-021-00500-1} {\bibfield  {journal} {\bibinfo
  {journal} {Light: Science \& Applications}\ }\textbf {\bibinfo {volume}
  {10}},\ \bibinfo {pages} {72} (\bibinfo {year} {2021})}\BibitemShut {NoStop}%
\bibitem [{\citenamefont {Greiner}\ \emph {et~al.}(2002)\citenamefont
  {Greiner}, \citenamefont {Mandel}, \citenamefont {Esslinger}, \citenamefont
  {Hänsch},\ and\ \citenamefont {Bloch}}]{greiner_quantum_2002}%
  \BibitemOpen
  \bibfield  {author} {\bibinfo {author} {\bibfnamefont {M.}~\bibnamefont
  {Greiner}}, \bibinfo {author} {\bibfnamefont {O.}~\bibnamefont {Mandel}},
  \bibinfo {author} {\bibfnamefont {T.}~\bibnamefont {Esslinger}}, \bibinfo
  {author} {\bibfnamefont {T.~W.}\ \bibnamefont {Hänsch}},\ and\ \bibinfo
  {author} {\bibfnamefont {I.}~\bibnamefont {Bloch}},\ }\bibfield  {title}
  {\bibinfo {title} {Quantum phase transition from a superfluid to a {Mott}
  insulator in a gas of ultracold atoms},\ }\href
  {https://doi.org/10.1038/415039a} {\bibfield  {journal} {\bibinfo  {journal}
  {Nature}\ }\textbf {\bibinfo {volume} {415}},\ \bibinfo {pages} {39}
  (\bibinfo {year} {2002})}\BibitemShut {NoStop}%
\bibitem [{\citenamefont {Jaksch}\ and\ \citenamefont
  {Zoller}(2005)}]{jaksch_cold_2005}%
  \BibitemOpen
  \bibfield  {author} {\bibinfo {author} {\bibfnamefont {D.}~\bibnamefont
  {Jaksch}}\ and\ \bibinfo {author} {\bibfnamefont {P.}~\bibnamefont
  {Zoller}},\ }\bibfield  {title} {\bibinfo {title} {The cold atom {Hubbard}
  toolbox},\ }\href {https://doi.org/10.1016/j.aop.2004.09.010} {\bibfield
  {journal} {\bibinfo  {journal} {Annals of Physics}\ }\textbf {\bibinfo
  {volume} {315}},\ \bibinfo {pages} {52} (\bibinfo {year} {2005})}\BibitemShut
  {NoStop}%
\bibitem [{\citenamefont {Lagoin}\ and\ \citenamefont
  {Dubin}(2021)}]{lagoin_key_2021}%
  \BibitemOpen
  \bibfield  {author} {\bibinfo {author} {\bibfnamefont {C.}~\bibnamefont
  {Lagoin}}\ and\ \bibinfo {author} {\bibfnamefont {F.}~\bibnamefont {Dubin}},\
  }\bibfield  {title} {\bibinfo {title} {Key role of the moiré potential for
  the quasicondensation of interlayer excitons in van der {Waals}
  heterostructures},\ }\href {https://doi.org/10.1103/PhysRevB.103.L041406}
  {\bibfield  {journal} {\bibinfo  {journal} {Physical Review B}\ }\textbf
  {\bibinfo {volume} {103}},\ \bibinfo {pages} {L041406} (\bibinfo {year}
  {2021})}\BibitemShut {NoStop}%
\bibitem [{\citenamefont {Wang}\ \emph {et~al.}(2021)\citenamefont {Wang},
  \citenamefont {Shi}, \citenamefont {Shih}, \citenamefont {Zhou},
  \citenamefont {Wu}, \citenamefont {Bai}, \citenamefont {Rhodes},
  \citenamefont {Barmak}, \citenamefont {Hone}, \citenamefont {Dean},\ and\
  \citenamefont {Zhu}}]{wang_diffusivity_2021}%
  \BibitemOpen
  \bibfield  {author} {\bibinfo {author} {\bibfnamefont {J.}~\bibnamefont
  {Wang}}, \bibinfo {author} {\bibfnamefont {Q.}~\bibnamefont {Shi}}, \bibinfo
  {author} {\bibfnamefont {E.-M.}\ \bibnamefont {Shih}}, \bibinfo {author}
  {\bibfnamefont {L.}~\bibnamefont {Zhou}}, \bibinfo {author} {\bibfnamefont
  {W.}~\bibnamefont {Wu}}, \bibinfo {author} {\bibfnamefont {Y.}~\bibnamefont
  {Bai}}, \bibinfo {author} {\bibfnamefont {D.}~\bibnamefont {Rhodes}},
  \bibinfo {author} {\bibfnamefont {K.}~\bibnamefont {Barmak}}, \bibinfo
  {author} {\bibfnamefont {J.}~\bibnamefont {Hone}}, \bibinfo {author}
  {\bibfnamefont {C.~R.}\ \bibnamefont {Dean}},\ and\ \bibinfo {author}
  {\bibfnamefont {X.-Y.}\ \bibnamefont {Zhu}},\ }\bibfield  {title} {\bibinfo
  {title} {Diffusivity reveals three distinct phases of interlayer excitons in
  {MoSe$_2$} / {WSe$_2$} heterobilayers},\ }\href
  {https://doi.org/10.1103/PhysRevLett.126.106804} {\bibfield  {journal}
  {\bibinfo  {journal} {Physical Review Letters}\ }\textbf {\bibinfo {volume}
  {126}},\ \bibinfo {pages} {106804} (\bibinfo {year} {2021})}\BibitemShut
  {NoStop}%
\bibitem [{\citenamefont {Lagoin}\ \emph {et~al.}(2020)\citenamefont {Lagoin},
  \citenamefont {Suffit}, \citenamefont {Bernard}, \citenamefont {Vabre},
  \citenamefont {West}, \citenamefont {Baldwin}, \citenamefont {Pfeiffer},\
  and\ \citenamefont {Dubin}}]{lagoin_microscopic_2020}%
  \BibitemOpen
  \bibfield  {author} {\bibinfo {author} {\bibfnamefont {C.}~\bibnamefont
  {Lagoin}}, \bibinfo {author} {\bibfnamefont {S.}~\bibnamefont {Suffit}},
  \bibinfo {author} {\bibfnamefont {M.}~\bibnamefont {Bernard}}, \bibinfo
  {author} {\bibfnamefont {M.}~\bibnamefont {Vabre}}, \bibinfo {author}
  {\bibfnamefont {K.}~\bibnamefont {West}}, \bibinfo {author} {\bibfnamefont
  {K.}~\bibnamefont {Baldwin}}, \bibinfo {author} {\bibfnamefont
  {L.}~\bibnamefont {Pfeiffer}},\ and\ \bibinfo {author} {\bibfnamefont
  {F.}~\bibnamefont {Dubin}},\ }\bibfield  {title} {\bibinfo {title}
  {Microscopic lattice for two-dimensional dipolar excitons},\ }\href
  {https://doi.org/10.1103/PhysRevB.102.245428} {\bibfield  {journal} {\bibinfo
   {journal} {Physical Review B}\ }\textbf {\bibinfo {volume} {102}},\ \bibinfo
  {pages} {245428} (\bibinfo {year} {2020})}\BibitemShut {NoStop}%
\bibitem [{\citenamefont {Lagoin}\ \emph {et~al.}(2021)\citenamefont {Lagoin},
  \citenamefont {Suffit}, \citenamefont {West}, \citenamefont {Baldwin},
  \citenamefont {Pfeiffer}, \citenamefont {Holzmann},\ and\ \citenamefont
  {Dubin}}]{lagoin_quasicondensation_2021}%
  \BibitemOpen
  \bibfield  {author} {\bibinfo {author} {\bibfnamefont {C.}~\bibnamefont
  {Lagoin}}, \bibinfo {author} {\bibfnamefont {S.}~\bibnamefont {Suffit}},
  \bibinfo {author} {\bibfnamefont {K.}~\bibnamefont {West}}, \bibinfo {author}
  {\bibfnamefont {K.}~\bibnamefont {Baldwin}}, \bibinfo {author} {\bibfnamefont
  {L.}~\bibnamefont {Pfeiffer}}, \bibinfo {author} {\bibfnamefont
  {M.}~\bibnamefont {Holzmann}},\ and\ \bibinfo {author} {\bibfnamefont
  {F.}~\bibnamefont {Dubin}},\ }\bibfield  {title} {\bibinfo {title}
  {Quasicondensation of bilayer excitons in a periodic potential},\ }\href
  {https://doi.org/10.1103/PhysRevLett.126.067404} {\bibfield  {journal}
  {\bibinfo  {journal} {Physical Review Letters}\ }\textbf {\bibinfo {volume}
  {126}},\ \bibinfo {pages} {067404} (\bibinfo {year} {2021})}\BibitemShut
  {NoStop}%
\bibitem [{\citenamefont {Steinke}\ \emph {et~al.}(2017)\citenamefont
  {Steinke}, \citenamefont {Mourad}, \citenamefont {Rösner}, \citenamefont
  {Lorke}, \citenamefont {Gies}, \citenamefont {Jahnke}, \citenamefont
  {Czycholl},\ and\ \citenamefont {Wehling}}]{steinke_noninvasive_2017}%
  \BibitemOpen
  \bibfield  {author} {\bibinfo {author} {\bibfnamefont {C.}~\bibnamefont
  {Steinke}}, \bibinfo {author} {\bibfnamefont {D.}~\bibnamefont {Mourad}},
  \bibinfo {author} {\bibfnamefont {M.}~\bibnamefont {Rösner}}, \bibinfo
  {author} {\bibfnamefont {M.}~\bibnamefont {Lorke}}, \bibinfo {author}
  {\bibfnamefont {C.}~\bibnamefont {Gies}}, \bibinfo {author} {\bibfnamefont
  {F.}~\bibnamefont {Jahnke}}, \bibinfo {author} {\bibfnamefont
  {G.}~\bibnamefont {Czycholl}},\ and\ \bibinfo {author} {\bibfnamefont
  {T.~O.}\ \bibnamefont {Wehling}},\ }\bibfield  {title} {\bibinfo {title}
  {Noninvasive control of excitons in two-dimensional materials},\ }\href
  {https://doi.org/10.1103/PhysRevB.96.045431} {\bibfield  {journal} {\bibinfo
  {journal} {Physical Review B}\ }\textbf {\bibinfo {volume} {96}},\ \bibinfo
  {pages} {045431} (\bibinfo {year} {2017})}\BibitemShut {NoStop}%
\bibitem [{\citenamefont {Borghardt}\ \emph {et~al.}(2017)\citenamefont
  {Borghardt}, \citenamefont {Tu}, \citenamefont {Winkler}, \citenamefont
  {Schubert}, \citenamefont {Zander}, \citenamefont {Leosson},\ and\
  \citenamefont {Kardynał}}]{borghardt_engineering_2017}%
  \BibitemOpen
  \bibfield  {author} {\bibinfo {author} {\bibfnamefont {S.}~\bibnamefont
  {Borghardt}}, \bibinfo {author} {\bibfnamefont {J.-S.}\ \bibnamefont {Tu}},
  \bibinfo {author} {\bibfnamefont {F.}~\bibnamefont {Winkler}}, \bibinfo
  {author} {\bibfnamefont {J.}~\bibnamefont {Schubert}}, \bibinfo {author}
  {\bibfnamefont {W.}~\bibnamefont {Zander}}, \bibinfo {author} {\bibfnamefont
  {K.}~\bibnamefont {Leosson}},\ and\ \bibinfo {author} {\bibfnamefont {B.~E.}\
  \bibnamefont {Kardynał}},\ }\bibfield  {title} {\bibinfo {title}
  {Engineering of optical and electronic band gaps in transition metal
  dichalcogenide monolayers through external dielectric screening},\ }\href
  {https://doi.org/10.1103/PhysRevMaterials.1.054001} {\bibfield  {journal}
  {\bibinfo  {journal} {Physical Review Materials}\ }\textbf {\bibinfo {volume}
  {1}},\ \bibinfo {pages} {054001} (\bibinfo {year} {2017})}\BibitemShut
  {NoStop}%
\bibitem [{\citenamefont {Kyl\"{a}np\"{a}\"{a}}\ and\ \citenamefont
  {Komsa}(2015)}]{kylanpaa_binding_2015}%
  \BibitemOpen
  \bibfield  {author} {\bibinfo {author} {\bibfnamefont {I.}~\bibnamefont
  {Kyl\"{a}np\"{a}\"{a}}}\ and\ \bibinfo {author} {\bibfnamefont {H.-P.}\
  \bibnamefont {Komsa}},\ }\bibfield  {title} {\bibinfo {title} {Binding
  energies of exciton complexes in transition metal dichalcogenide monolayers
  and effect of dielectric environment},\ }\href
  {https://doi.org/10.1103/PhysRevB.92.205418} {\bibfield  {journal} {\bibinfo
  {journal} {Physical Review B}\ }\textbf {\bibinfo {volume} {92}},\ \bibinfo
  {pages} {205418} (\bibinfo {year} {2015})}\BibitemShut {NoStop}%
\bibitem [{\citenamefont {Berman}\ and\ \citenamefont
  {Kezerashvili}(2017)}]{berman_superfluidity_2017}%
  \BibitemOpen
  \bibfield  {author} {\bibinfo {author} {\bibfnamefont {O.~L.}\ \bibnamefont
  {Berman}}\ and\ \bibinfo {author} {\bibfnamefont {R.~Y.}\ \bibnamefont
  {Kezerashvili}},\ }\bibfield  {title} {\bibinfo {title} {Superfluidity of
  dipolar excitons in a transition metal dichalcogenide double layer},\ }\href
  {https://doi.org/10.1103/PhysRevB.96.094502} {\bibfield  {journal} {\bibinfo
  {journal} {Physical Review B}\ }\textbf {\bibinfo {volume} {96}},\ \bibinfo
  {pages} {094502} (\bibinfo {year} {2017})}\BibitemShut {NoStop}%
\bibitem [{\citenamefont {Rytova}(1967)}]{rytova1967the8248}%
  \BibitemOpen
  \bibfield  {author} {\bibinfo {author} {\bibfnamefont {N.~S.}\ \bibnamefont
  {Rytova}},\ }\bibfield  {title} {\bibinfo {title} {The screened potential of
  a point charge in a thin film},\ }\href@noop {} {\bibfield  {journal}
  {\bibinfo  {journal} {Moscow Univ.\ Phys.\ Bull.}\ }\textbf {\bibinfo
  {volume} {3}},\ \bibinfo {pages} {18} (\bibinfo {year} {1967})}\BibitemShut
  {NoStop}%
\bibitem [{\citenamefont {Keldysh}(1979)}]{keldysh_coulomb_1979}%
  \BibitemOpen
  \bibfield  {author} {\bibinfo {author} {\bibfnamefont {L.~V.}\ \bibnamefont
  {Keldysh}},\ }\bibfield  {title} {\bibinfo {title} {Coulomb interactions in
  thin semiconductor and semimetal films},\ }\href@noop {} {\bibfield
  {journal} {\bibinfo  {journal} {JETP Lett.}\ }\textbf {\bibinfo {volume}
  {29}},\ \bibinfo {pages} {658} (\bibinfo {year} {1979})}\BibitemShut
  {NoStop}%
\bibitem [{\citenamefont {Cudazzo}\ \emph {et~al.}(2011)\citenamefont
  {Cudazzo}, \citenamefont {Tokatly},\ and\ \citenamefont
  {Rubio}}]{cudazzo_dielectric_2011}%
  \BibitemOpen
  \bibfield  {author} {\bibinfo {author} {\bibfnamefont {P.}~\bibnamefont
  {Cudazzo}}, \bibinfo {author} {\bibfnamefont {I.~V.}\ \bibnamefont
  {Tokatly}},\ and\ \bibinfo {author} {\bibfnamefont {A.}~\bibnamefont
  {Rubio}},\ }\bibfield  {title} {\bibinfo {title} {Dielectric screening in
  two-dimensional insulators: {Implications} for excitonic and impurity states
  in graphane},\ }\href {https://doi.org/10.1103/PhysRevB.84.085406} {\bibfield
   {journal} {\bibinfo  {journal} {Physical Review B}\ }\textbf {\bibinfo
  {volume} {84}},\ \bibinfo {pages} {085406} (\bibinfo {year}
  {2011})}\BibitemShut {NoStop}%
\bibitem [{\citenamefont {Rodin}\ \emph {et~al.}(2014)\citenamefont {Rodin},
  \citenamefont {Carvalho},\ and\ \citenamefont
  {Castro~Neto}}]{rodin_excitons_2014}%
  \BibitemOpen
  \bibfield  {author} {\bibinfo {author} {\bibfnamefont {A.~S.}\ \bibnamefont
  {Rodin}}, \bibinfo {author} {\bibfnamefont {A.}~\bibnamefont {Carvalho}},\
  and\ \bibinfo {author} {\bibfnamefont {A.~H.}\ \bibnamefont {Castro~Neto}},\
  }\bibfield  {title} {\bibinfo {title} {Excitons in anisotropic
  two-dimensional semiconducting crystals},\ }\href
  {https://doi.org/10.1103/PhysRevB.90.075429} {\bibfield  {journal} {\bibinfo
  {journal} {Physical Review B}\ }\textbf {\bibinfo {volume} {90}},\ \bibinfo
  {pages} {075429} (\bibinfo {year} {2014})}\BibitemShut {NoStop}%
\bibitem [{\citenamefont {Terrones}\ \emph {et~al.}(2013)\citenamefont
  {Terrones}, \citenamefont {López-Urias},\ and\ \citenamefont
  {Terrones}}]{terrones_novel_2013}%
  \BibitemOpen
  \bibfield  {author} {\bibinfo {author} {\bibfnamefont {H.}~\bibnamefont
  {Terrones}}, \bibinfo {author} {\bibfnamefont {F.}~\bibnamefont
  {López-Urias}},\ and\ \bibinfo {author} {\bibfnamefont {M.}~\bibnamefont
  {Terrones}},\ }\bibfield  {title} {\bibinfo {title} {Novel hetero-layered
  materials with tunable direct band gaps by sandwiching different metal
  disulfides and diselenides},\ }\href {https://doi.org/10.1038/srep01549}
  {\bibfield  {journal} {\bibinfo  {journal} {Scientific Reports}\ }\textbf
  {\bibinfo {volume} {3}},\ \bibinfo {pages} {1549} (\bibinfo {year}
  {2013})}\BibitemShut {NoStop}%
\bibitem [{\citenamefont {Hong}\ \emph {et~al.}(2014)\citenamefont {Hong},
  \citenamefont {Kim}, \citenamefont {Shi}, \citenamefont {Zhang},
  \citenamefont {Jin}, \citenamefont {Sun}, \citenamefont {Tongay},
  \citenamefont {Wu}, \citenamefont {Zhang},\ and\ \citenamefont
  {Wang}}]{hong_ultrafast_2014}%
  \BibitemOpen
  \bibfield  {author} {\bibinfo {author} {\bibfnamefont {X.}~\bibnamefont
  {Hong}}, \bibinfo {author} {\bibfnamefont {J.}~\bibnamefont {Kim}}, \bibinfo
  {author} {\bibfnamefont {S.-F.}\ \bibnamefont {Shi}}, \bibinfo {author}
  {\bibfnamefont {Y.}~\bibnamefont {Zhang}}, \bibinfo {author} {\bibfnamefont
  {C.}~\bibnamefont {Jin}}, \bibinfo {author} {\bibfnamefont {Y.}~\bibnamefont
  {Sun}}, \bibinfo {author} {\bibfnamefont {S.}~\bibnamefont {Tongay}},
  \bibinfo {author} {\bibfnamefont {J.}~\bibnamefont {Wu}}, \bibinfo {author}
  {\bibfnamefont {Y.}~\bibnamefont {Zhang}},\ and\ \bibinfo {author}
  {\bibfnamefont {F.}~\bibnamefont {Wang}},\ }\bibfield  {title} {\bibinfo
  {title} {Ultrafast charge transfer in atomically thin {MoS$_2$}/{WS$_2$}
  heterostructures},\ }\href {https://doi.org/10.1038/nnano.2014.167}
  {\bibfield  {journal} {\bibinfo  {journal} {Nature Nanotechnology}\ }\textbf
  {\bibinfo {volume} {9}},\ \bibinfo {pages} {682} (\bibinfo {year}
  {2014})}\BibitemShut {NoStop}%
\bibitem [{\citenamefont {Chen}\ \emph {et~al.}(2016)\citenamefont {Chen},
  \citenamefont {Wen}, \citenamefont {Zhang}, \citenamefont {Wu}, \citenamefont
  {Gong}, \citenamefont {Zhang}, \citenamefont {Yuan}, \citenamefont {Yi},
  \citenamefont {Lou}, \citenamefont {Ajayan}, \citenamefont {Zhuang},
  \citenamefont {Zhang},\ and\ \citenamefont {Zheng}}]{chen_ultrafast_2016}%
  \BibitemOpen
  \bibfield  {author} {\bibinfo {author} {\bibfnamefont {H.}~\bibnamefont
  {Chen}}, \bibinfo {author} {\bibfnamefont {X.}~\bibnamefont {Wen}}, \bibinfo
  {author} {\bibfnamefont {J.}~\bibnamefont {Zhang}}, \bibinfo {author}
  {\bibfnamefont {T.}~\bibnamefont {Wu}}, \bibinfo {author} {\bibfnamefont
  {Y.}~\bibnamefont {Gong}}, \bibinfo {author} {\bibfnamefont {X.}~\bibnamefont
  {Zhang}}, \bibinfo {author} {\bibfnamefont {J.}~\bibnamefont {Yuan}},
  \bibinfo {author} {\bibfnamefont {C.}~\bibnamefont {Yi}}, \bibinfo {author}
  {\bibfnamefont {J.}~\bibnamefont {Lou}}, \bibinfo {author} {\bibfnamefont
  {P.~M.}\ \bibnamefont {Ajayan}}, \bibinfo {author} {\bibfnamefont
  {W.}~\bibnamefont {Zhuang}}, \bibinfo {author} {\bibfnamefont
  {G.}~\bibnamefont {Zhang}},\ and\ \bibinfo {author} {\bibfnamefont
  {J.}~\bibnamefont {Zheng}},\ }\bibfield  {title} {\bibinfo {title} {Ultrafast
  formation of interlayer hot excitons in atomically thin {MoS$_2$}/{WS$_2$}
  heterostructures},\ }\href {https://doi.org/10.1038/ncomms12512} {\bibfield
  {journal} {\bibinfo  {journal} {Nature Communications}\ }\textbf {\bibinfo
  {volume} {7}},\ \bibinfo {pages} {12512} (\bibinfo {year}
  {2016})}\BibitemShut {NoStop}%
\bibitem [{\citenamefont {Jin}\ \emph {et~al.}(2018)\citenamefont {Jin},
  \citenamefont {Ma}, \citenamefont {Karni}, \citenamefont {Regan},
  \citenamefont {Wang},\ and\ \citenamefont {Heinz}}]{jin_ultrafast_2018}%
  \BibitemOpen
  \bibfield  {author} {\bibinfo {author} {\bibfnamefont {C.}~\bibnamefont
  {Jin}}, \bibinfo {author} {\bibfnamefont {E.~Y.}\ \bibnamefont {Ma}},
  \bibinfo {author} {\bibfnamefont {O.}~\bibnamefont {Karni}}, \bibinfo
  {author} {\bibfnamefont {E.~C.}\ \bibnamefont {Regan}}, \bibinfo {author}
  {\bibfnamefont {F.}~\bibnamefont {Wang}},\ and\ \bibinfo {author}
  {\bibfnamefont {T.~F.}\ \bibnamefont {Heinz}},\ }\bibfield  {title} {\bibinfo
  {title} {Ultrafast dynamics in van der {Waals} heterostructures},\ }\href
  {https://doi.org/10.1038/s41565-018-0298-5} {\bibfield  {journal} {\bibinfo
  {journal} {Nature Nanotechnology}\ }\textbf {\bibinfo {volume} {13}},\
  \bibinfo {pages} {994} (\bibinfo {year} {2018})}\BibitemShut {NoStop}%
\bibitem [{\citenamefont {Choi}\ \emph {et~al.}(2021)\citenamefont {Choi},
  \citenamefont {Florian}, \citenamefont {Steinhoff}, \citenamefont {Erben},
  \citenamefont {Tran}, \citenamefont {Kim}, \citenamefont {Sun}, \citenamefont
  {Quan}, \citenamefont {Claassen}, \citenamefont {Majumder}, \citenamefont
  {Hollingsworth}, \citenamefont {Taniguchi}, \citenamefont {Watanabe},
  \citenamefont {Ueno}, \citenamefont {Singh}, \citenamefont {Moody},
  \citenamefont {Jahnke},\ and\ \citenamefont {Li}}]{choi_twist_2021}%
  \BibitemOpen
  \bibfield  {author} {\bibinfo {author} {\bibfnamefont {J.}~\bibnamefont
  {Choi}}, \bibinfo {author} {\bibfnamefont {M.}~\bibnamefont {Florian}},
  \bibinfo {author} {\bibfnamefont {A.}~\bibnamefont {Steinhoff}}, \bibinfo
  {author} {\bibfnamefont {D.}~\bibnamefont {Erben}}, \bibinfo {author}
  {\bibfnamefont {K.}~\bibnamefont {Tran}}, \bibinfo {author} {\bibfnamefont
  {D.~S.}\ \bibnamefont {Kim}}, \bibinfo {author} {\bibfnamefont
  {L.}~\bibnamefont {Sun}}, \bibinfo {author} {\bibfnamefont {J.}~\bibnamefont
  {Quan}}, \bibinfo {author} {\bibfnamefont {R.}~\bibnamefont {Claassen}},
  \bibinfo {author} {\bibfnamefont {S.}~\bibnamefont {Majumder}}, \bibinfo
  {author} {\bibfnamefont {J.~A.}\ \bibnamefont {Hollingsworth}}, \bibinfo
  {author} {\bibfnamefont {T.}~\bibnamefont {Taniguchi}}, \bibinfo {author}
  {\bibfnamefont {K.}~\bibnamefont {Watanabe}}, \bibinfo {author}
  {\bibfnamefont {K.}~\bibnamefont {Ueno}}, \bibinfo {author} {\bibfnamefont
  {A.}~\bibnamefont {Singh}}, \bibinfo {author} {\bibfnamefont
  {G.}~\bibnamefont {Moody}}, \bibinfo {author} {\bibfnamefont
  {F.}~\bibnamefont {Jahnke}},\ and\ \bibinfo {author} {\bibfnamefont
  {X.}~\bibnamefont {Li}},\ }\bibfield  {title} {\bibinfo {title} {Twist
  {Angle}-{Dependent} {Interlayer} {Exciton} {Lifetimes} in van der {Waals}
  {Heterostructures}},\ }\href {https://doi.org/10.1103/PhysRevLett.126.047401}
  {\bibfield  {journal} {\bibinfo  {journal} {Physical Review Letters}\
  }\textbf {\bibinfo {volume} {126}},\ \bibinfo {pages} {047401} (\bibinfo
  {year} {2021})}\BibitemShut {NoStop}%
\bibitem [{\citenamefont {Zimmermann}\ and\ \citenamefont
  {Schindler}(2007)}]{zimmermann_excitonexciton_2007}%
  \BibitemOpen
  \bibfield  {author} {\bibinfo {author} {\bibfnamefont {R.}~\bibnamefont
  {Zimmermann}}\ and\ \bibinfo {author} {\bibfnamefont {C.}~\bibnamefont
  {Schindler}},\ }\bibfield  {title} {\bibinfo {title} {Exciton–exciton
  interaction in coupled quantum wells},\ }\href
  {https://doi.org/10.1016/j.ssc.2007.07.044} {\bibfield  {journal} {\bibinfo
  {journal} {Solid State Communications}\ }\textbf {\bibinfo {volume} {144}},\
  \bibinfo {pages} {395} (\bibinfo {year} {2007})}\BibitemShut {NoStop}%
\bibitem [{\citenamefont {Schindler}\ and\ \citenamefont
  {Zimmermann}(2008)}]{schindler_analysis_2008}%
  \BibitemOpen
  \bibfield  {author} {\bibinfo {author} {\bibfnamefont {C.}~\bibnamefont
  {Schindler}}\ and\ \bibinfo {author} {\bibfnamefont {R.}~\bibnamefont
  {Zimmermann}},\ }\bibfield  {title} {\bibinfo {title} {Analysis of the
  exciton-exciton interaction in semiconductor quantum wells},\ }\href
  {https://doi.org/10.1103/PhysRevB.78.045313} {\bibfield  {journal} {\bibinfo
  {journal} {Physical Review B}\ }\textbf {\bibinfo {volume} {78}},\ \bibinfo
  {pages} {045313} (\bibinfo {year} {2008})}\BibitemShut {NoStop}%
\bibitem [{\citenamefont {Laikhtman}\ and\ \citenamefont
  {Rapaport}(2009)}]{laikhtman_exciton_2009}%
  \BibitemOpen
  \bibfield  {author} {\bibinfo {author} {\bibfnamefont {B.}~\bibnamefont
  {Laikhtman}}\ and\ \bibinfo {author} {\bibfnamefont {R.}~\bibnamefont
  {Rapaport}},\ }\bibfield  {title} {\bibinfo {title} {Exciton correlations in
  coupled quantum wells and their luminescence blue shift},\ }\href
  {https://doi.org/10.1103/PhysRevB.80.195313} {\bibfield  {journal} {\bibinfo
  {journal} {Physical Review B}\ }\textbf {\bibinfo {volume} {80}},\ \bibinfo
  {pages} {195313} (\bibinfo {year} {2009})}\BibitemShut {NoStop}%
\bibitem [{\citenamefont {Batsch}\ \emph {et~al.}(1993)\citenamefont {Batsch},
  \citenamefont {Meier}, \citenamefont {Thomas}, \citenamefont {Lindberg},
  \citenamefont {Koch},\ and\ \citenamefont
  {Shah}}]{batsch_dipole-dipole_1993}%
  \BibitemOpen
  \bibfield  {author} {\bibinfo {author} {\bibfnamefont {M.}~\bibnamefont
  {Batsch}}, \bibinfo {author} {\bibfnamefont {T.}~\bibnamefont {Meier}},
  \bibinfo {author} {\bibfnamefont {P.}~\bibnamefont {Thomas}}, \bibinfo
  {author} {\bibfnamefont {M.}~\bibnamefont {Lindberg}}, \bibinfo {author}
  {\bibfnamefont {S.~W.}\ \bibnamefont {Koch}},\ and\ \bibinfo {author}
  {\bibfnamefont {J.}~\bibnamefont {Shah}},\ }\bibfield  {title} {\bibinfo
  {title} {Dipole-dipole coupling of excitons in double quantum wells},\ }\href
  {https://doi.org/10.1103/PhysRevB.48.11817} {\bibfield  {journal} {\bibinfo
  {journal} {Physical Review B}\ }\textbf {\bibinfo {volume} {48}},\ \bibinfo
  {pages} {11817} (\bibinfo {year} {1993})}\BibitemShut {NoStop}%
\bibitem [{\citenamefont {Li}\ \emph {et~al.}(2020)\citenamefont {Li},
  \citenamefont {Lu}, \citenamefont {Dubey}, \citenamefont {Devenica},\ and\
  \citenamefont {Srivastava}}]{li_dipolar_2020}%
  \BibitemOpen
  \bibfield  {author} {\bibinfo {author} {\bibfnamefont {W.}~\bibnamefont
  {Li}}, \bibinfo {author} {\bibfnamefont {X.}~\bibnamefont {Lu}}, \bibinfo
  {author} {\bibfnamefont {S.}~\bibnamefont {Dubey}}, \bibinfo {author}
  {\bibfnamefont {L.}~\bibnamefont {Devenica}},\ and\ \bibinfo {author}
  {\bibfnamefont {A.}~\bibnamefont {Srivastava}},\ }\bibfield  {title}
  {\bibinfo {title} {Dipolar interactions between localized interlayer excitons
  in van der {Waals} heterostructures},\ }\href
  {https://doi.org/10.1038/s41563-020-0661-4} {\bibfield  {journal} {\bibinfo
  {journal} {Nature Materials}\ }\textbf {\bibinfo {volume} {19}},\ \bibinfo
  {pages} {624} (\bibinfo {year} {2020})}\BibitemShut {NoStop}%
\bibitem [{\citenamefont {Laturia}\ \emph {et~al.}(2018)\citenamefont
  {Laturia}, \citenamefont {Van~de Put},\ and\ \citenamefont
  {Vandenberghe}}]{laturia_dielectric_2018}%
  \BibitemOpen
  \bibfield  {author} {\bibinfo {author} {\bibfnamefont {A.}~\bibnamefont
  {Laturia}}, \bibinfo {author} {\bibfnamefont {M.~L.}\ \bibnamefont {Van~de
  Put}},\ and\ \bibinfo {author} {\bibfnamefont {W.~G.}\ \bibnamefont
  {Vandenberghe}},\ }\bibfield  {title} {\bibinfo {title} {Dielectric
  properties of hexagonal boron nitride and transition metal dichalcogenides:
  from monolayer to bulk},\ }\href {https://doi.org/10.1038/s41699-018-0050-x}
  {\bibfield  {journal} {\bibinfo  {journal} {npj 2D Materials and
  Applications}\ }\textbf {\bibinfo {volume} {2}},\ \bibinfo {pages} {6}
  (\bibinfo {year} {2018})}\BibitemShut {NoStop}%
\bibitem [{\citenamefont {Thygesen}(2017)}]{thygesen_calculating_2017}%
  \BibitemOpen
  \bibfield  {author} {\bibinfo {author} {\bibfnamefont {K.~S.}\ \bibnamefont
  {Thygesen}},\ }\bibfield  {title} {\bibinfo {title} {Calculating excitons,
  plasmons, and quasiparticles in {2D} materials and van der {Waals}
  heterostructures},\ }\href {https://doi.org/10.1088/2053-1583/aa6432}
  {\bibfield  {journal} {\bibinfo  {journal} {2D Materials}\ }\textbf {\bibinfo
  {volume} {4}},\ \bibinfo {pages} {022004} (\bibinfo {year}
  {2017})}\BibitemShut {NoStop}%
\bibitem [{\citenamefont {Velický}\ and\ \citenamefont
  {Toth}(2017)}]{velicky_two-dimensional_2017}%
  \BibitemOpen
  \bibfield  {author} {\bibinfo {author} {\bibfnamefont {M.}~\bibnamefont
  {Velický}}\ and\ \bibinfo {author} {\bibfnamefont {P.~S.}\ \bibnamefont
  {Toth}},\ }\bibfield  {title} {\bibinfo {title} {From two-dimensional
  materials to their heterostructures: {An} electrochemist's perspective},\
  }\href {https://doi.org/10.1016/j.apmt.2017.05.003} {\bibfield  {journal}
  {\bibinfo  {journal} {Applied Materials Today}\ }\textbf {\bibinfo {volume}
  {8}},\ \bibinfo {pages} {68} (\bibinfo {year} {2017})}\BibitemShut {NoStop}%
\bibitem [{\citenamefont {Florian}\ \emph {et~al.}(2018)\citenamefont
  {Florian}, \citenamefont {Hartmann}, \citenamefont {Steinhoff}, \citenamefont
  {Klein}, \citenamefont {Holleitner}, \citenamefont {Finley}, \citenamefont
  {Wehling}, \citenamefont {Kaniber},\ and\ \citenamefont
  {Gies}}]{florian_dielectric_2018}%
  \BibitemOpen
  \bibfield  {author} {\bibinfo {author} {\bibfnamefont {M.}~\bibnamefont
  {Florian}}, \bibinfo {author} {\bibfnamefont {M.}~\bibnamefont {Hartmann}},
  \bibinfo {author} {\bibfnamefont {A.}~\bibnamefont {Steinhoff}}, \bibinfo
  {author} {\bibfnamefont {J.}~\bibnamefont {Klein}}, \bibinfo {author}
  {\bibfnamefont {A.~W.}\ \bibnamefont {Holleitner}}, \bibinfo {author}
  {\bibfnamefont {J.~J.}\ \bibnamefont {Finley}}, \bibinfo {author}
  {\bibfnamefont {T.~O.}\ \bibnamefont {Wehling}}, \bibinfo {author}
  {\bibfnamefont {M.}~\bibnamefont {Kaniber}},\ and\ \bibinfo {author}
  {\bibfnamefont {C.}~\bibnamefont {Gies}},\ }\bibfield  {title} {\bibinfo
  {title} {The dielectric impact of layer distances on exciton and trion
  binding energies in van der {Waals} heterostructures},\ }\href
  {https://doi.org/10.1021/acs.nanolett.8b00840} {\bibfield  {journal}
  {\bibinfo  {journal} {Nano Letters}\ }\textbf {\bibinfo {volume} {18}},\
  \bibinfo {pages} {2725} (\bibinfo {year} {2018})}\BibitemShut {NoStop}%
\bibitem [{\citenamefont {Van~Tuan}\ \emph {et~al.}(2018)\citenamefont
  {Van~Tuan}, \citenamefont {Yang},\ and\ \citenamefont
  {Dery}}]{van_tuan_coulomb_2018}%
  \BibitemOpen
  \bibfield  {author} {\bibinfo {author} {\bibfnamefont {D.}~\bibnamefont
  {Van~Tuan}}, \bibinfo {author} {\bibfnamefont {M.}~\bibnamefont {Yang}},\
  and\ \bibinfo {author} {\bibfnamefont {H.}~\bibnamefont {Dery}},\ }\bibfield
  {title} {\bibinfo {title} {Coulomb interaction in monolayer transition-metal
  dichalcogenides},\ }\href {https://doi.org/10.1103/PhysRevB.98.125308}
  {\bibfield  {journal} {\bibinfo  {journal} {Physical Review B}\ }\textbf
  {\bibinfo {volume} {98}},\ \bibinfo {pages} {125308} (\bibinfo {year}
  {2018})}\BibitemShut {NoStop}%
\bibitem [{\citenamefont {Berkelbach}\ \emph {et~al.}(2013)\citenamefont
  {Berkelbach}, \citenamefont {Hybertsen},\ and\ \citenamefont
  {Reichman}}]{berkelbach_theory_2013}%
  \BibitemOpen
  \bibfield  {author} {\bibinfo {author} {\bibfnamefont {T.~C.}\ \bibnamefont
  {Berkelbach}}, \bibinfo {author} {\bibfnamefont {M.~S.}\ \bibnamefont
  {Hybertsen}},\ and\ \bibinfo {author} {\bibfnamefont {D.~R.}\ \bibnamefont
  {Reichman}},\ }\bibfield  {title} {\bibinfo {title} {Theory of neutral and
  charged excitons in monolayer transition metal dichalcogenides},\ }\href
  {https://doi.org/10.1103/PhysRevB.88.045318} {\bibfield  {journal} {\bibinfo
  {journal} {Physical Review B}\ }\textbf {\bibinfo {volume} {88}},\ \bibinfo
  {pages} {045318} (\bibinfo {year} {2013})}\BibitemShut {NoStop}%
\bibitem [{\citenamefont {Berghäuser}\ and\ \citenamefont
  {Malic}(2014)}]{berghauser_analytical_2014}%
  \BibitemOpen
  \bibfield  {author} {\bibinfo {author} {\bibfnamefont {G.}~\bibnamefont
  {Berghäuser}}\ and\ \bibinfo {author} {\bibfnamefont {E.}~\bibnamefont
  {Malic}},\ }\bibfield  {title} {\bibinfo {title} {Analytical approach to
  excitonic properties of {MoS$_2$}},\ }\href
  {https://doi.org/10.1103/PhysRevB.89.125309} {\bibfield  {journal} {\bibinfo
  {journal} {Physical Review B}\ }\textbf {\bibinfo {volume} {89}},\ \bibinfo
  {pages} {125309} (\bibinfo {year} {2014})}\BibitemShut {NoStop}%
\bibitem [{\citenamefont {Capogrosso-Sansone}\ \emph
  {et~al.}(2008)\citenamefont {Capogrosso-Sansone}, \citenamefont {S\"{o}yler},
  \citenamefont {Prokof'ev},\ and\ \citenamefont
  {Svistunov}}]{capogrosso-sansone_monte_2008}%
  \BibitemOpen
  \bibfield  {author} {\bibinfo {author} {\bibfnamefont {B.}~\bibnamefont
  {Capogrosso-Sansone}}, \bibinfo {author} {\bibfnamefont {S.~G.}\ \bibnamefont
  {S\"{o}yler}}, \bibinfo {author} {\bibfnamefont {N.}~\bibnamefont
  {Prokof'ev}},\ and\ \bibinfo {author} {\bibfnamefont {B.}~\bibnamefont
  {Svistunov}},\ }\bibfield  {title} {\bibinfo {title} {Monte {Carlo} study of
  the two-dimensional {Bose}-{Hubbard} model},\ }\href
  {https://doi.org/10.1103/PhysRevA.77.015602} {\bibfield  {journal} {\bibinfo
  {journal} {Physical Review A}\ }\textbf {\bibinfo {volume} {77}},\ \bibinfo
  {pages} {015602} (\bibinfo {year} {2008})}\BibitemShut {NoStop}%
\bibitem [{\citenamefont {Bogner}\ \emph {et~al.}(2019)\citenamefont {Bogner},
  \citenamefont {De~Daniloff},\ and\ \citenamefont
  {Rieger}}]{bogner_variational_2019}%
  \BibitemOpen
  \bibfield  {author} {\bibinfo {author} {\bibfnamefont {B.}~\bibnamefont
  {Bogner}}, \bibinfo {author} {\bibfnamefont {C.}~\bibnamefont
  {De~Daniloff}},\ and\ \bibinfo {author} {\bibfnamefont {H.}~\bibnamefont
  {Rieger}},\ }\bibfield  {title} {\bibinfo {title} {Variational
  {Monte}-{Carlo} study of the extended {Bose}-{Hubbard} model with short- and
  infinite-range interactions},\ }\href
  {https://doi.org/10.1140/epjb/e2019-100017-8} {\bibfield  {journal} {\bibinfo
   {journal} {The European Physical Journal B}\ }\textbf {\bibinfo {volume}
  {92}},\ \bibinfo {pages} {111} (\bibinfo {year} {2019})}\BibitemShut
  {NoStop}%
\bibitem [{\citenamefont {Kato}\ \emph {et~al.}(2007)\citenamefont {Kato},
  \citenamefont {Suzuki},\ and\ \citenamefont
  {Kawashima}}]{kato_modification_2007}%
  \BibitemOpen
  \bibfield  {author} {\bibinfo {author} {\bibfnamefont {Y.}~\bibnamefont
  {Kato}}, \bibinfo {author} {\bibfnamefont {T.}~\bibnamefont {Suzuki}},\ and\
  \bibinfo {author} {\bibfnamefont {N.}~\bibnamefont {Kawashima}},\ }\bibfield
  {title} {\bibinfo {title} {Modification of directed-loop algorithm for
  continuous space simulation of bosonic systems},\ }\href
  {https://doi.org/10.1103/PhysRevE.75.066703} {\bibfield  {journal} {\bibinfo
  {journal} {Physical Review E}\ }\textbf {\bibinfo {volume} {75}},\ \bibinfo
  {pages} {066703} (\bibinfo {year} {2007})}\BibitemShut {NoStop}%
\bibitem [{\citenamefont {Polkovnikov}(2010)}]{polkovnikov_phase_2010}%
  \BibitemOpen
  \bibfield  {author} {\bibinfo {author} {\bibfnamefont {A.}~\bibnamefont
  {Polkovnikov}},\ }\bibfield  {title} {\bibinfo {title} {Phase space
  representation of quantum dynamics},\ }\href
  {https://doi.org/10.1016/j.aop.2010.02.006} {\bibfield  {journal} {\bibinfo
  {journal} {Annals of Physics}\ }\textbf {\bibinfo {volume} {325}},\ \bibinfo
  {pages} {1790} (\bibinfo {year} {2010})}\BibitemShut {NoStop}%
\bibitem [{\citenamefont {Weiss}\ \emph {et~al.}(2018)\citenamefont {Weiss},
  \citenamefont {Gerster}, \citenamefont {Jaschke}, \citenamefont {Silvi},\
  and\ \citenamefont {Montangero}}]{weiss_kibble-zurek_2018}%
  \BibitemOpen
  \bibfield  {author} {\bibinfo {author} {\bibfnamefont {W.}~\bibnamefont
  {Weiss}}, \bibinfo {author} {\bibfnamefont {M.}~\bibnamefont {Gerster}},
  \bibinfo {author} {\bibfnamefont {D.}~\bibnamefont {Jaschke}}, \bibinfo
  {author} {\bibfnamefont {P.}~\bibnamefont {Silvi}},\ and\ \bibinfo {author}
  {\bibfnamefont {S.}~\bibnamefont {Montangero}},\ }\bibfield  {title}
  {\bibinfo {title} {Kibble-{Zurek} scaling of the one-dimensional
  {Bose}-{Hubbard} model at finite temperatures},\ }\href
  {https://doi.org/10.1103/PhysRevA.98.063601} {\bibfield  {journal} {\bibinfo
  {journal} {Physical Review A}\ }\textbf {\bibinfo {volume} {98}},\ \bibinfo
  {pages} {063601} (\bibinfo {year} {2018})}\BibitemShut {NoStop}%
\bibitem [{\citenamefont {Iblisdir}\ \emph {et~al.}(2007)\citenamefont
  {Iblisdir}, \citenamefont {Orús},\ and\ \citenamefont
  {Latorre}}]{iblisdir_matrix_2007}%
  \BibitemOpen
  \bibfield  {author} {\bibinfo {author} {\bibfnamefont {S.}~\bibnamefont
  {Iblisdir}}, \bibinfo {author} {\bibfnamefont {R.}~\bibnamefont {Orús}},\
  and\ \bibinfo {author} {\bibfnamefont {J.~I.}\ \bibnamefont {Latorre}},\
  }\bibfield  {title} {\bibinfo {title} {Matrix product states algorithms and
  continuous systems},\ }\href {https://doi.org/10.1103/PhysRevB.75.104305}
  {\bibfield  {journal} {\bibinfo  {journal} {Physical Review B}\ }\textbf
  {\bibinfo {volume} {75}},\ \bibinfo {pages} {104305} (\bibinfo {year}
  {2007})}\BibitemShut {NoStop}%
\bibitem [{\citenamefont {Garcia-Ripoll}\ \emph {et~al.}(2004)\citenamefont
  {Garcia-Ripoll}, \citenamefont {Cirac}, \citenamefont {Zoller}, \citenamefont
  {Kollath}, \citenamefont {Schollw\"{o}ck},\ and\ \citenamefont {von
  Delft}}]{garca-ripoll_variational_2004}%
  \BibitemOpen
  \bibfield  {author} {\bibinfo {author} {\bibfnamefont {J.~J.}\ \bibnamefont
  {Garcia-Ripoll}}, \bibinfo {author} {\bibfnamefont {J.~I.}\ \bibnamefont
  {Cirac}}, \bibinfo {author} {\bibfnamefont {P.}~\bibnamefont {Zoller}},
  \bibinfo {author} {\bibfnamefont {C.}~\bibnamefont {Kollath}}, \bibinfo
  {author} {\bibfnamefont {U.}~\bibnamefont {Schollw\"{o}ck}},\ and\ \bibinfo
  {author} {\bibfnamefont {J.}~\bibnamefont {von Delft}},\ }\bibfield  {title}
  {\bibinfo {title} {Variational ansatz for the superfluid {Mott}-insulator
  transition in optical lattices},\ }\href
  {https://doi.org/10.1364/OPEX.12.000042} {\bibfield  {journal} {\bibinfo
  {journal} {Optics Express}\ }\textbf {\bibinfo {volume} {12}},\ \bibinfo
  {pages} {42} (\bibinfo {year} {2004})}\BibitemShut {NoStop}%
\bibitem [{\citenamefont {Gygi}\ \emph {et~al.}(2006)\citenamefont {Gygi},
  \citenamefont {Katzgraber}, \citenamefont {Troyer}, \citenamefont {Wessel},\
  and\ \citenamefont {Batrouni}}]{gygi_simulations_2006}%
  \BibitemOpen
  \bibfield  {author} {\bibinfo {author} {\bibfnamefont {O.}~\bibnamefont
  {Gygi}}, \bibinfo {author} {\bibfnamefont {H.~G.}\ \bibnamefont
  {Katzgraber}}, \bibinfo {author} {\bibfnamefont {M.}~\bibnamefont {Troyer}},
  \bibinfo {author} {\bibfnamefont {S.}~\bibnamefont {Wessel}},\ and\ \bibinfo
  {author} {\bibfnamefont {G.~G.}\ \bibnamefont {Batrouni}},\ }\bibfield
  {title} {\bibinfo {title} {Simulations of ultracold bosonic atoms in optical
  lattices with anharmonic traps},\ }\href
  {https://doi.org/10.1103/PhysRevA.73.063606} {\bibfield  {journal} {\bibinfo
  {journal} {Physical Review A}\ }\textbf {\bibinfo {volume} {73}},\ \bibinfo
  {pages} {063606} (\bibinfo {year} {2006})}\BibitemShut {NoStop}%
\bibitem [{\citenamefont {Ohgoe}\ \emph {et~al.}(2012)\citenamefont {Ohgoe},
  \citenamefont {Suzuki},\ and\ \citenamefont
  {Kawashima}}]{ohgoe_ground-state_2012}%
  \BibitemOpen
  \bibfield  {author} {\bibinfo {author} {\bibfnamefont {T.}~\bibnamefont
  {Ohgoe}}, \bibinfo {author} {\bibfnamefont {T.}~\bibnamefont {Suzuki}},\ and\
  \bibinfo {author} {\bibfnamefont {N.}~\bibnamefont {Kawashima}},\ }\bibfield
  {title} {\bibinfo {title} {Ground-state phase diagram of the two-dimensional
  extended {Bose}-{Hubbard} model},\ }\href
  {https://doi.org/10.1103/PhysRevB.86.054520} {\bibfield  {journal} {\bibinfo
  {journal} {Physical Review B}\ }\textbf {\bibinfo {volume} {86}},\ \bibinfo
  {pages} {054520} (\bibinfo {year} {2012})}\BibitemShut {NoStop}%
\bibitem [{\citenamefont {Xu}\ \emph {et~al.}(2020)\citenamefont {Xu},
  \citenamefont {Liu}, \citenamefont {Rhodes}, \citenamefont {Watanabe},
  \citenamefont {Taniguchi}, \citenamefont {Hone}, \citenamefont {Elser},
  \citenamefont {Mak},\ and\ \citenamefont {Shan}}]{xu_correlated_2020}%
  \BibitemOpen
  \bibfield  {author} {\bibinfo {author} {\bibfnamefont {Y.}~\bibnamefont
  {Xu}}, \bibinfo {author} {\bibfnamefont {S.}~\bibnamefont {Liu}}, \bibinfo
  {author} {\bibfnamefont {D.~A.}\ \bibnamefont {Rhodes}}, \bibinfo {author}
  {\bibfnamefont {K.}~\bibnamefont {Watanabe}}, \bibinfo {author}
  {\bibfnamefont {T.}~\bibnamefont {Taniguchi}}, \bibinfo {author}
  {\bibfnamefont {J.}~\bibnamefont {Hone}}, \bibinfo {author} {\bibfnamefont
  {V.}~\bibnamefont {Elser}}, \bibinfo {author} {\bibfnamefont {K.~F.}\
  \bibnamefont {Mak}},\ and\ \bibinfo {author} {\bibfnamefont {J.}~\bibnamefont
  {Shan}},\ }\bibfield  {title} {\bibinfo {title} {Correlated insulating states
  at fractional fillings of moiré superlattices},\ }\href
  {https://doi.org/10.1038/s41586-020-2868-6} {\bibfield  {journal} {\bibinfo
  {journal} {Nature}\ }\textbf {\bibinfo {volume} {587}},\ \bibinfo {pages}
  {214} (\bibinfo {year} {2020})}\BibitemShut {NoStop}%
\bibitem [{\citenamefont {Gu}\ \emph {et~al.}(2022)\citenamefont {Gu},
  \citenamefont {Ma}, \citenamefont {Liu}, \citenamefont {Watanabe},
  \citenamefont {Taniguchi}, \citenamefont {Hone}, \citenamefont {Shan},\ and\
  \citenamefont {Mak}}]{gu_dipolar_2022}%
  \BibitemOpen
  \bibfield  {author} {\bibinfo {author} {\bibfnamefont {J.}~\bibnamefont
  {Gu}}, \bibinfo {author} {\bibfnamefont {L.}~\bibnamefont {Ma}}, \bibinfo
  {author} {\bibfnamefont {S.}~\bibnamefont {Liu}}, \bibinfo {author}
  {\bibfnamefont {K.}~\bibnamefont {Watanabe}}, \bibinfo {author}
  {\bibfnamefont {T.}~\bibnamefont {Taniguchi}}, \bibinfo {author}
  {\bibfnamefont {J.~C.}\ \bibnamefont {Hone}}, \bibinfo {author}
  {\bibfnamefont {J.}~\bibnamefont {Shan}},\ and\ \bibinfo {author}
  {\bibfnamefont {K.~F.}\ \bibnamefont {Mak}},\ }\bibfield  {title} {\bibinfo
  {title} {Dipolar excitonic insulator in a moiré lattice},\ }\href
  {https://doi.org/10.1038/s41567-022-01532-z} {\bibfield  {journal} {\bibinfo
  {journal} {Nature Physics}\ ,\ \bibinfo {pages} {1}} (\bibinfo {year}
  {2022})}\BibitemShut {NoStop}%
\bibitem [{\citenamefont {Regan}\ \emph {et~al.}(2020)\citenamefont {Regan},
  \citenamefont {Wang}, \citenamefont {Jin}, \citenamefont {Bakti~Utama},
  \citenamefont {Gao}, \citenamefont {Wei}, \citenamefont {Zhao}, \citenamefont
  {Zhao}, \citenamefont {Zhang}, \citenamefont {Yumigeta}, \citenamefont
  {Blei}, \citenamefont {Carlström}, \citenamefont {Watanabe}, \citenamefont
  {Taniguchi}, \citenamefont {Tongay}, \citenamefont {Crommie}, \citenamefont
  {Zettl},\ and\ \citenamefont {Wang}}]{regan_mott_2020}%
  \BibitemOpen
  \bibfield  {author} {\bibinfo {author} {\bibfnamefont {E.~C.}\ \bibnamefont
  {Regan}}, \bibinfo {author} {\bibfnamefont {D.}~\bibnamefont {Wang}},
  \bibinfo {author} {\bibfnamefont {C.}~\bibnamefont {Jin}}, \bibinfo {author}
  {\bibfnamefont {M.~I.}\ \bibnamefont {Bakti~Utama}}, \bibinfo {author}
  {\bibfnamefont {B.}~\bibnamefont {Gao}}, \bibinfo {author} {\bibfnamefont
  {X.}~\bibnamefont {Wei}}, \bibinfo {author} {\bibfnamefont {S.}~\bibnamefont
  {Zhao}}, \bibinfo {author} {\bibfnamefont {W.}~\bibnamefont {Zhao}}, \bibinfo
  {author} {\bibfnamefont {Z.}~\bibnamefont {Zhang}}, \bibinfo {author}
  {\bibfnamefont {K.}~\bibnamefont {Yumigeta}}, \bibinfo {author}
  {\bibfnamefont {M.}~\bibnamefont {Blei}}, \bibinfo {author} {\bibfnamefont
  {J.~D.}\ \bibnamefont {Carlström}}, \bibinfo {author} {\bibfnamefont
  {K.}~\bibnamefont {Watanabe}}, \bibinfo {author} {\bibfnamefont
  {T.}~\bibnamefont {Taniguchi}}, \bibinfo {author} {\bibfnamefont
  {S.}~\bibnamefont {Tongay}}, \bibinfo {author} {\bibfnamefont
  {M.}~\bibnamefont {Crommie}}, \bibinfo {author} {\bibfnamefont
  {A.}~\bibnamefont {Zettl}},\ and\ \bibinfo {author} {\bibfnamefont
  {F.}~\bibnamefont {Wang}},\ }\bibfield  {title} {\bibinfo {title} {Mott and
  generalized {Wigner} crystal states in {WSe$_2$}/{WS$_2$} moiré
  superlattices},\ }\href {https://doi.org/10.1038/s41586-020-2092-4}
  {\bibfield  {journal} {\bibinfo  {journal} {Nature}\ }\textbf {\bibinfo
  {volume} {579}},\ \bibinfo {pages} {359} (\bibinfo {year}
  {2020})}\BibitemShut {NoStop}%
\bibitem [{\citenamefont {Miao}\ \emph {et~al.}(2021)\citenamefont {Miao},
  \citenamefont {Wang}, \citenamefont {Huang}, \citenamefont {Chen},
  \citenamefont {Lian}, \citenamefont {Wang}, \citenamefont {Blei},
  \citenamefont {Taniguchi}, \citenamefont {Watanabe}, \citenamefont {Tongay},
  \citenamefont {Wang}, \citenamefont {Xiao}, \citenamefont {Cui},\ and\
  \citenamefont {Shi}}]{miao_strong_2021}%
  \BibitemOpen
  \bibfield  {author} {\bibinfo {author} {\bibfnamefont {S.}~\bibnamefont
  {Miao}}, \bibinfo {author} {\bibfnamefont {T.}~\bibnamefont {Wang}}, \bibinfo
  {author} {\bibfnamefont {X.}~\bibnamefont {Huang}}, \bibinfo {author}
  {\bibfnamefont {D.}~\bibnamefont {Chen}}, \bibinfo {author} {\bibfnamefont
  {Z.}~\bibnamefont {Lian}}, \bibinfo {author} {\bibfnamefont {C.}~\bibnamefont
  {Wang}}, \bibinfo {author} {\bibfnamefont {M.}~\bibnamefont {Blei}}, \bibinfo
  {author} {\bibfnamefont {T.}~\bibnamefont {Taniguchi}}, \bibinfo {author}
  {\bibfnamefont {K.}~\bibnamefont {Watanabe}}, \bibinfo {author}
  {\bibfnamefont {S.}~\bibnamefont {Tongay}}, \bibinfo {author} {\bibfnamefont
  {Z.}~\bibnamefont {Wang}}, \bibinfo {author} {\bibfnamefont {D.}~\bibnamefont
  {Xiao}}, \bibinfo {author} {\bibfnamefont {Y.-T.}\ \bibnamefont {Cui}},\ and\
  \bibinfo {author} {\bibfnamefont {S.-F.}\ \bibnamefont {Shi}},\ }\bibfield
  {title} {\bibinfo {title} {Strong interaction between interlayer excitons and
  correlated electrons in {WSe$_2$}/{WS$_2$} moiré superlattice},\ }\href
  {https://doi.org/10.1038/s41467-021-23732-6} {\bibfield  {journal} {\bibinfo
  {journal} {Nature Communications}\ }\textbf {\bibinfo {volume} {12}},\
  \bibinfo {pages} {3608} (\bibinfo {year} {2021})}\BibitemShut {NoStop}%
\bibitem [{\citenamefont {Huber}\ and\ \citenamefont
  {Altman}(2010)}]{huber_bose_2010}%
  \BibitemOpen
  \bibfield  {author} {\bibinfo {author} {\bibfnamefont {S.~D.}\ \bibnamefont
  {Huber}}\ and\ \bibinfo {author} {\bibfnamefont {E.}~\bibnamefont {Altman}},\
  }\bibfield  {title} {\bibinfo {title} {Bose condensation in flat bands},\
  }\href {https://doi.org/10.1103/PhysRevB.82.184502} {\bibfield  {journal}
  {\bibinfo  {journal} {Physical Review B}\ }\textbf {\bibinfo {volume} {82}},\
  \bibinfo {pages} {184502} (\bibinfo {year} {2010})}\BibitemShut {NoStop}%
\bibitem [{\citenamefont {Hui}\ \emph {et~al.}(2017)\citenamefont {Hui},
  \citenamefont {Zhang}, \citenamefont {Zhang},\ and\ \citenamefont
  {Scarola}}]{hui_superfluidity_2017}%
  \BibitemOpen
  \bibfield  {author} {\bibinfo {author} {\bibfnamefont {H.-Y.}\ \bibnamefont
  {Hui}}, \bibinfo {author} {\bibfnamefont {Y.}~\bibnamefont {Zhang}}, \bibinfo
  {author} {\bibfnamefont {C.}~\bibnamefont {Zhang}},\ and\ \bibinfo {author}
  {\bibfnamefont {V.~W.}\ \bibnamefont {Scarola}},\ }\bibfield  {title}
  {\bibinfo {title} {Superfluidity in the absence of kinetics in
  spin-orbit-coupled optical lattices},\ }\href
  {https://doi.org/10.1103/PhysRevA.95.033603} {\bibfield  {journal} {\bibinfo
  {journal} {Physical Review A}\ }\textbf {\bibinfo {volume} {95}},\ \bibinfo
  {pages} {033603} (\bibinfo {year} {2017})}\BibitemShut {NoStop}%
\bibitem [{\citenamefont {Zhao}\ \emph {et~al.}(2022)\citenamefont {Zhao},
  \citenamefont {Huang}, \citenamefont {Li}, \citenamefont {Rupp},
  \citenamefont {Göser}, \citenamefont {Vovk}, \citenamefont {Kruchinin},
  \citenamefont {Watanabe}, \citenamefont {Taniguchi}, \citenamefont {Bilgin},
  \citenamefont {Baimuratov},\ and\ \citenamefont
  {Högele}}]{zhao_excitons_2022}%
  \BibitemOpen
  \bibfield  {author} {\bibinfo {author} {\bibfnamefont {S.}~\bibnamefont
  {Zhao}}, \bibinfo {author} {\bibfnamefont {X.}~\bibnamefont {Huang}},
  \bibinfo {author} {\bibfnamefont {Z.}~\bibnamefont {Li}}, \bibinfo {author}
  {\bibfnamefont {A.}~\bibnamefont {Rupp}}, \bibinfo {author} {\bibfnamefont
  {J.}~\bibnamefont {Göser}}, \bibinfo {author} {\bibfnamefont {I.~A.}\
  \bibnamefont {Vovk}}, \bibinfo {author} {\bibfnamefont {S.~Y.}\ \bibnamefont
  {Kruchinin}}, \bibinfo {author} {\bibfnamefont {K.}~\bibnamefont {Watanabe}},
  \bibinfo {author} {\bibfnamefont {T.}~\bibnamefont {Taniguchi}}, \bibinfo
  {author} {\bibfnamefont {I.}~\bibnamefont {Bilgin}}, \bibinfo {author}
  {\bibfnamefont {A.~S.}\ \bibnamefont {Baimuratov}},\ and\ \bibinfo {author}
  {\bibfnamefont {A.}~\bibnamefont {Högele}},\ }\bibfield  {title} {\bibinfo
  {title} {Excitons in mesoscopically reconstructed moir\'{e}
  heterostructures},\ }\href {http://arxiv.org/abs/2202.11139} {\bibfield
  {journal} {\bibinfo  {journal} {arXiv:2202.11139}\ } (\bibinfo {year}
  {2022})}\BibitemShut {NoStop}%
\end{thebibliography}%

\end{document}